\documentclass[11pt,a4paper]{article}
\pdfoutput=1
\usepackage{jstyle}
\usepackage{amsmath, amsfonts, amssymb}
\usepackage{graphicx, hyperref, color, verbatim}
\usepackage[dvipsnames]{xcolor}
\usepackage{float}
\usepackage{caption}
\usepackage{subcaption}
\usepackage[bb=stix]{mathalpha}
\usepackage{tikz}
\usetikzlibrary{shapes}
\usetikzlibrary{shapes.misc}

\tikzset{
  vtx/.style={
    circle,
    draw=blue,
    fill=blue,
    inner sep=1pt
  },
  wcirc/.style={
    circle,
    draw=white,
    fill=white,
    inner sep=2pt
  },
  bcirc/.style={
    circle,
    draw=black,
    fill=black,
    inner sep=1pt
  },
  dcirc/.style={
    circle,
    draw=blue,
    fill=blue,
    inner sep=1pt
  },
  rcirc/.style={
    circle,
    draw=red,
    fill=red,
    inner sep=1pt
  },
  phi/.style={
    thick
  },
  sigma/.style={
    thick,
    dashed
  },
  vl1/.style={
    thick,
    blue
  },
  vl2/.style={
    thick,
    dashed,
    blue
  },
  valign/.style={
    baseline={([yshift=-.55ex]current bounding box.center)}
  }
}

\usepackage[LGR,T1]{fontenc}
\DeclareSymbolFont{ttgreek}{LGR}{cmtt}{m}{n}
\DeclareMathSymbol{\ttalpha}{\mathord}{ttgreek}{`a}

\makeatletter
\DeclareFontFamily{OMX}{MnSymbolE}{}
\DeclareSymbolFont{MnLargeSymbols}{OMX}{MnSymbolE}{m}{n}
\SetSymbolFont{MnLargeSymbols}{bold}{OMX}{MnSymbolE}{b}{n}
\DeclareFontShape{OMX}{MnSymbolE}{m}{n}{
    <-6>  MnSymbolE5
   <6-7>  MnSymbolE6
   <7-8>  MnSymbolE7
   <8-9>  MnSymbolE8
   <9-10> MnSymbolE9
  <10-12> MnSymbolE10
  <12->   MnSymbolE12
}{}
\DeclareFontShape{OMX}{MnSymbolE}{b}{n}{
    <-6>  MnSymbolE-Bold5
   <6-7>  MnSymbolE-Bold6
   <7-8>  MnSymbolE-Bold7
   <8-9>  MnSymbolE-Bold8
   <9-10> MnSymbolE-Bold9
  <10-12> MnSymbolE-Bold10
  <12->   MnSymbolE-Bold12
}{}

\let\llangle\@undefined
\let\rrangle\@undefined
\DeclareMathDelimiter{\llangle}{\mathopen}%
                     {MnLargeSymbols}{'164}{MnLargeSymbols}{'164}
\DeclareMathDelimiter{\rrangle}{\mathclose}%
                     {MnLargeSymbols}{'171}{MnLargeSymbols}{'171}
\makeatother

\def\M{{\mathcal{M}}}

\def\H{{\mathcal{H}}}
\def\P{\mathcal{P}}
\def\Q{{\mathcal{Q}}}

\def\xb{\bar{x}}
\def\zb{{\bar{z}}}

\def\Li{{\text{Li}}}
\def\dfourhat{{\widehat{\Delta}^{(4)}}}

\makeatletter
\def\@fpheader{\hfill USTC-ICTS/PCFT-26-49}
\makeatother

\title{Quantum Gravity Corrections to Giant Graviton Correlators}

\author[a,b]{Junding Chen,}
\author[a,b]{Xi-Er Du,}
\author[c]{Hynek Paul,}
\author[b,d]{Xinan Zhou}

\affiliation[a]{School of Physical Sciences, University of Chinese Academy of Sciences, No.19A Yuquan Road, Beijing 100049, China}
\affiliation[b]{Kavli Institute for Theoretical Sciences, University of Chinese Academy of Sciences, Beijing 100190, China}
\affiliation[c]{Instituut voor Theoretische Fysica, KU Leuven, Celestijnenlaan 200D, B-3001 Leuven, Belgium}
\affiliation[d]{Peng Huanwu Center for Fundamental Theory, Hefei, Anhui 230026, China}

\abstract{We compute the leading quantum gravity correction to the correlation function of two maximal giant gravitons and two light supergravitons in the stress tensor multiplet in the supergravity limit of $\mathcal{N}=4$ SYM theory. By viewing the giant gravitons as a zero-dimensional defect, we employ the defect version of the AdS unitarity method to reconstruct the one-loop correction from the known tree-level correlators. We obtain the one-loop result in closed form in both Mellin and position space. We also extract the quantum correction to the anomalous dimension of operators describing a bound state of a giant graviton and a light supergraviton at the leading conformal twist in the defect channel OPE.}

\begin{document}
\maketitle
\tableofcontents

\newpage

\section{Introduction}
The planar limit of $\mathcal{N}=4$ SYM theory is well under control thanks to several complementary approaches. At small 't Hooft coupling, standard perturbation theory is further simplified by the restriction to planar Feynman diagrams. When the coupling is large, the theory admits a dual description via AdS/CFT as classical supergravity in AdS$_5\times$S$^5$. In addition, integrability provides another handle which works even at finite coupling. Meanwhile, it should be noted that the planar regime is the place where most systematic progress has been achieved. Moreover, in many cases this is restricted to light probes, i.e., single-trace operators whose dimensions do not scale with $N$. Going away from the planar limit, eventually to finite $N$, turns out to be very difficult as these tools which we heavily rely on become either ineffective or inefficient.

Correlation functions of giant graviton operators \cite{McGreevy:2000cw,Balasubramanian:2001nh,Hashimoto:2000zp} provide a good intermediate target that sits  between the usual planar limit and finite $N$. Giant gravitons are $\frac{1}{2}$-BPS operators and therefore preserve a large amount of supersymmetry. Their conformal dimensions scale linearly with $N$, meaning they are not light excitations as the single-trace operators. Therefore, correlators involving giant gravitons strike the right balance:  supersymmetric enough to have sufficient control, but at the same time heavy enough to retain genuinely non-planar effects even when $N$ is taken to infinity. Recently, significant progress has been made in computing giant graviton correlators at strong coupling. In \cite{Chen:2025yxg, Chen:2026ium}, all four-point functions of two maximal giant gravitons and two light supergravitons of arbitrary dimensions have been computed in the large $N$ regime corresponding to tree-level AdS supergravity.\footnote{See also \cite{Bissi:2011dc,Caputa:2012yj,Jiang:2019xdz,Jiang:2019zig,Holguin:2022zii} for progress on three-point functions, and \cite{Jiang:2019xdz,Jiang:2019zig,Jiang:2023uut,He:2026ios} for four-point functions at weak coupling.} Two important ingredients are responsible for this development. The first is a novel defect perspective which views the giant graviton insertions as a special kind of zero-dimensional defect \cite{Chen:2025yxg, Chen:2026ium}. More generally, the defect perspective has proven to be a natural language for light-light-heavy-heavy correlators that can be turned into a concrete computational framework. The second is the bootstrap strategy for computing holographic correlators, which was initiated in \cite{Rastelli:2016nze,Rastelli:2017udc} and recently adapted to defects \cite{Gimenez-Grau:2023fcy,Chen:2023yvw,Zhou:2024ekb}. This strategy allows one to efficiently compute defect correlators without using explicit details of the complicated effective action. In this work, we continue to use this approach and build on the results of \cite{Chen:2025yxg, Chen:2026ium} by considering the leading quantum gravity correction to the maximal giant graviton correlators.

Our main technical tool for computing the one-loop quantum correction is the defect version of the AdS unitarity method, which schematically glues the tree-level correlators of \cite{Chen:2025yxg, Chen:2026ium} into the one-loop correlator. The defect unitarity method has already been developed in \cite{Chen:2024orp} where it was applied to surface defects in 6d $(2,0)$ theories. Here we apply it to the case of giant gravitons, which amounts to restricting the defect dimension to zero. In addition to demonstrating the generality of the method in a different setup, we also extend it in an important way. In \cite{Chen:2024orp} the unitarity method was only developed in Mellin space, and developing a position space counterpart was mentioned as an open problem. In this work, we close this gap by presenting a complementary position space computation of the one-loop correlator based on the same unitarity principle. In particular, we provide a concrete answer to the question of what constitutes the basis of functions in position space. The main findings of our paper are summarized as follows.
\begin{itemize}
    \item {\bf Mellin space.} We apply the method of \cite{Chen:2024orp} and obtain the one-loop giant graviton correlator in Mellin space in a closed form. The result for the Mellin amplitude is structurally similar to the 6d case: the amplitude also contains only simultaneous poles and the numerator coefficients can be expressed in terms of the same type of hypergeometric functions.  
    \item {\bf Position space.} We also perform a complementary position-space bootstrap similar to the light operator four-point function case  \cite{Aprile:2017bgs,Aprile:2017qoy,Aprile:2019rep,Huang:2021xws,Drummond:2022dxw,Huang:2023oxf}. We find that the one-loop correlator can be expressed in terms of similar set of transcendental functions, albeit of maximal transcendental degree 3. This allows us to write down a position-space ansatz with finitely many unknowns. Imposing a set of physical conditions similar to the defect-free case fixes most of the coefficients but still leaves a few undetermined. These remaining coefficients can be fixed by comparing against the Mellin space answer. The final result can be written in a compact form by using a fourth-order differential operator introduced in \cite{Chen:2026ium}.

    \item {\bf Flat-space limit.} We also study the flat-space limit of the one-loop correction as a nontrivial consistency check. We find that the reduced Mellin amplitude in this limit simply becomes a single one-loop Feynman diagram in flat space which involves massless particles scattering off an extended defect.

    \item {\bf Anomalous dimensions.} In the defect channel OPE, we find operators built from a giant graviton and a light supergravity mode. Such operators have anomalous dimensions which correspond to the binding energies of these bound states in AdS. From the one-loop corrected correlator, we extract the quantum gravity correction to the binding energies for operators with the leading defect conformal twist. 
\end{itemize}
The rest of the paper is organized as follows. In Section \ref{sec:kinematics}, we review the basic kinematics of the giant graviton correlators. In Section \ref{section:unitarity_method}, we review the AdS unitarity method for holographic defects and compute the leading logarithmic singularities. We then compute the full one-loop correlator in Mellin space in Section \ref{Sec:Mellinspace}, where we also study the flat-space limit. In Section \ref{Sec:positionspace}, we compute the one-loop correlator  in position space and also extract the quantum correction to the anomalous dimensions. We conclude in Section \ref{Sec:discussion} by mentioning several open questions and future directions.

\section{Kinematics}\label{sec:kinematics}

\noindent{\bf Operators and correlators.} We consider correlation functions of light supergravitons and heavy giant gravitons. The supergravitons are $\frac{1}{2}$-BPS operators of the form 
\begin{equation}
    \mathcal{O}_k(x, t) = \text{tr}(X^I t^I)^k+\ldots\;,
\end{equation}
where $X^I$ are the six scalars of $\mathcal{N}=4$ SYM and $t^I$ is a null R-symmetry polarization vector satisfying $t\cdot t=0$. The $\ldots$ stand for $1/N$ suppressed multi-trace corrections so that $\mathcal{O}_k$ corresponds to a single-particle Kaluza-Klein mode in AdS. The operator has protected conformal dimension $k$ and transforms in the rank-$k$ symmetric traceless representation of $SO(6)$. The giant graviton operator takes the form 
\begin{equation}
    \mathcal{D}(x, t) = \det(X\cdot t)\;,
\end{equation}
and is also a $\frac{1}{2}$-BPS operator, with dimension $N$ and transforms in the rank-$N$ symmetric traceless representation. In the bulk, the giant graviton operator corresponds to a D3 brane which wraps an S$^3\subset~$S$^5$. Our focus is the four-point function
\begin{equation}
    \langle \mathcal{O}_{k_1}(x_1,t_1) \mathcal{O}_{k_2}(x_2,t_2) \mathcal{D}(x_3,t_3) \mathcal{D}(x_4,t_4) \rangle\;.
\end{equation}

\vspace{0.5cm}

\noindent{\bf Superconformal constraints.} The superconformal kinematics of such $\frac{1}{2}$-BPS four-point correlators is well understood in the literature \cite{Eden:2000bk,Nirschl:2004pa}. However, in the large-$N$ limit the giant graviton operators becomes infinitely heavy and it was pointed out in \cite{Chen:2025yxg,Chen:2026ium} that it is more advantageous to view these correlators as two-point functions in the presence of a zero-dimensional defect. We define the defect correlator as the ratio  
\begin{equation}
    G_{k_1k_2}(x_i,t_i)=\llangle \mathcal{O}_{k_1}(x_1,t_1) \mathcal{O}_{k_2}(x_2,t_2) \rrangle = \frac{\langle \mathcal{O}_{k_1} (x_1,t_1)\mathcal{O}_{k_2} (x_2,t_2)\mathcal{D} (x_3,t_3) \mathcal{D} (x_4,t_4) \rangle}{\langle \mathcal{D} (x_3,t_3) \mathcal{D} (x_4,t_4) \rangle} \,,
\end{equation}
where the quantum numbers of the heavy operators become invisible. The s-channel OPE of the four-point function should be identified with the bulk channel of the defect two-point function, while the t- and u-channels should be identified with the defect channel. It is convenient to further extract a factor of defect one-point functions to write the correlator as a function of cross ratios
\begin{equation}
   G_{k_1k_2}(x_i,t_i)=\left( \prod_{i=1,2} \frac{t_{i3}t_{i4}x_{34}^2}{t_{34}x_{i3}^2 x_{i4}^2} \right)^{\frac{k_i}{2}} \mathcal{G}_{k_1k_2}(U,V,\sigma,\tau)\;,
\end{equation}
where
\begin{equation}
     U = \frac{x_{12}^2 x_{34}^2}{x_{13}^2 x_{24}^2} \,, \quad V = \frac{x_{14}^2 x_{23}^2}{x_{13}^2 x_{24}^2}\;,\quad \sigma = \frac{t_{13}t_{24}}{t_{12}t_{34}} \,, \quad \tau = \frac{t_{14}t_{23}}{t_{12}t_{34}}\;,
\end{equation}
with $x_{ij}=x_i-x_j$ and $t_{ij}=t_i\cdot t_j$. By further taking into account the constraints from the fermionic symmetry generators, which can be achieved by using the known four-point function results of \cite{Eden:2000bk,Nirschl:2004pa}, we can write the defect two-point function as  
\begin{equation}\label{solscfWardid}
    \mathcal{G}_{k_1k_2}=\mathcal{G}_{k_1k_2,{\rm free}}+(V\sigma\tau)^{-1}R\,\mathcal{H}_{k_1k_2}\;.
\end{equation}
Here $\mathcal{G}_{k_1k_2,{\rm free}}$ is the defect two-point function in the free theory and all the dynamical information is captured by the reduced correlator $\mathcal{H}_{k_1k_2}$. The factor $R$ is fixed by superconformal symmetry to be
\begin{equation}
    R=(1-z\alpha)(1-z\bar{\alpha})(1-\bar{z}\alpha)(1-\bar{z}\bar{\alpha})\;,
\end{equation}  
where we introduced the change of variables
\begin{equation}\label{eq:cross_ratios}
    U=z\bar{z}\;,\quad V=(1-z)(1-\bar{z})\;,\quad \sigma=\alpha\bar{\alpha}\;,\quad \tau=(1-\alpha)(1-\bar{\alpha})\;.
\end{equation}
We can also restore the kinematic factors extracted in $\mathcal{H}_{k_1k_2}$
\begin{equation}\label{reducedcorrelator}
 H_{k_1k_2}=   \left( \prod_{i=1,2} \frac{t_{i3}t_{i4}x_{34}^2}{t_{34}x_{i3}^2 x_{i4}^2} \right)^{\frac{k_i+2}{2}}\frac{t_{34}^4}{t_{13}^2t_{14}^2t_{23}^2t_{24}^2}\,\mathcal{H}_{k_1k_2}\;.
\end{equation}
This manifests the fact that the reduced correlator $H_{k_1k_2}$ has shifted conformal dimensions $\{k_1+2,k_2+2\}$ and R-symmetry charges $\{k_1-2,k_2-2\}$.

\vspace{0.5cm}

\noindent{\bf Large-$c$ expansion.} In this paper, we consider the giant graviton correlators in the supergravity limit where the central charge $c=\frac{1}{4}(N^2-1)$ is large. The correlators can be expanded into inverse powers of $c$ as
\begin{equation}\label{eq:G_large-c}
    G_{k_1k_2}=\left(G^{(0)}_{\rm bulk}+G^{(0)}_{\rm defect}\right)+\frac{1}{\sqrt{c}}G^{(1)}_{\rm tree}+\frac{1}{c}\left(G^{(2)}_{\rm 1-loop}+G^{(1)}_{\rm disc}\right)+\ldots\;,
\end{equation}
where each order can be matched with the diagrammatic expansion in AdS in a way similar to \cite{Chen:2024orp}. The leading-order term in the large-$c$ expansion has two contributions:
\begin{equation}\label{eq:G0}
    G^{(0)}_{\rm bulk}=\langle \mathcal{O}_{k_1}\mathcal{O}_{k_2}\rangle \delta_{k_1k_2}\;, \quad G^{(0)}_{\rm defect}=\llangle \mathcal{O}_{k_1}\rrangle \llangle \mathcal{O}_{k_2}\rrangle\;.
\end{equation}
The first term is a free propagator in AdS and is only present when the light operators are identical. The second term is a product of one-point functions. These two terms correspond to the exchange of the identity operator in the bulk and defect channel, respectively. At the subleading $\mathcal{O}(c^{-1/2})$ order, the correlators are given by sums of tree-level Witten diagrams 
\begin{align}
G^{(1)}_{\rm tree}
 \;\;=\;\;
   \begin{tikzpicture}[valign,scale=0.8]
    \tikzstyle{every node}=[font=\scriptsize]
    \pgfmathsetmacro{\x}{-sqrt(2)/2}
    \pgfmathsetmacro{\y}{sqrt(2)/2}
    \pgfmathsetmacro{\z}{0.5}
    \pgfmathsetmacro{\r}{0.2}
    \draw [thick] (0,0) circle [radius=1];
    \draw [thick, blue] (-\x,-\y) to[out=150,in=30] (+\x,-\y);
    \draw  (+\x,+\y) -- (0,\r);
    \draw  (-\x,+\y) -- (0,\r);
    \draw (0,\r) -- (0,-\z);
     \fill [red!80!black] (+\x,+\y) circle (2.5pt);
     \fill [red!80!black] (-\x,+\y) circle (2.5pt);
    \fill [black] (0, \r) circle (2.5pt);
    \fill [green!60!black] (0,-\z) circle (2.5pt);
 \end{tikzpicture}
 \;\;+\;\;
 \begin{tikzpicture}[valign,scale=0.8]
    \tikzstyle{every node}=[font=\scriptsize]
    \pgfmathsetmacro{\xx}{sqrt(2)/2}
    \pgfmathsetmacro{\yy}{sqrt(2)/2}
    \pgfmathsetmacro{\z}{0.5}
    \pgfmathsetmacro{\rx}{0.32}
    \pgfmathsetmacro{\ry}{0.54}
    \draw [thick] (0,0) circle [radius=1];
    \draw      (-\xx,+\yy) -- (-\rx,-\ry);
    \draw      (+\xx,+\yy) -- (\rx,-\ry);
    \draw [densely dashed,blue] (-\rx,-\ry) to[out=100,in=80] (\rx,-\ry);
    \draw [thick, blue] (\xx,-\yy) to[out=150,in=30] (-\xx,-\yy);
     \fill [red!80!black] (-\xx,+\yy) circle (2.5pt);
     \fill [red!80!black] (+\xx,+\yy) circle (2.5pt);
    \fill [green!60!black] (-\rx,-\ry) circle (2.5pt);
    \fill [green!60!black] (\rx,-\ry) circle (2.5pt);
 \end{tikzpicture}
 \;\;+\;\;
 \begin{tikzpicture}[valign,scale=0.8]
    \tikzstyle{every node}=[font=\scriptsize]
    \pgfmathsetmacro{\x}{-sqrt(2)/2}
    \pgfmathsetmacro{\y}{sqrt(2)/2}
    \pgfmathsetmacro{\z}{0.5}
    \pgfmathsetmacro{\r}{-0.75}
    \draw [thick] (0,0) circle [radius=1];
    \draw [thick, blue] (-\x,-\y) to[out=150,in=30] (+\x,-\y);
    \draw     (-\x,+\y) -- (0,-\z);
    \draw     (+\x,+\y) -- (0,-\z);
       \fill [red!80!black]  (-\x,+\y) circle (2.5pt);
     \fill [red!80!black] (+\x,+\y) circle (2.5pt);
    \fill [green!60!black] (0,-\z) circle (2.5pt);
  \end{tikzpicture}~,
\end{align}
and they have been computed in \cite{Chen:2025yxg,Chen:2026ium}. The focus of this paper is the next order $\mathcal{O}(c^{-1})$. We are interested in the contributions $G^{(2)}_{\rm 1-loop}$ coming from one-loop level Witten diagrams
\begin{align}
 G^{(2)}_{\rm 1-loop}
 \;\;=\;\;
   \begin{tikzpicture}[valign,scale=0.8]
    \tikzstyle{every node}=[font=\scriptsize]
    \pgfmathsetmacro{\x}{-sqrt(2)/2}
    \pgfmathsetmacro{\y}{sqrt(2)/2}
    \pgfmathsetmacro{\z}{0.2}
    \pgfmathsetmacro{\r}{0.2}
    \pgfmathsetmacro{\xx}{0.4}
    \pgfmathsetmacro{\yy}{0.55}
    \draw [thick] (0,0) circle [radius=1];
    \draw [thick, blue] (-\x,-\y) to[out=150,in=30] (+\x,-\y);
    \draw  (+\x,+\y) -- (0,\r);
    \draw  (-\x,+\y) -- (0,\r);
    \draw (0,\r) -- (0,-\z);
    \draw (-\xx,-\yy) -- (0,-\z);
    \draw (\xx,-\yy) -- (0,-\z);
     \fill [red!80!black] (+\x,+\y) circle (2.5pt);
     \fill [red!80!black] (-\x,+\y) circle (2.5pt);
    \fill [black] (0, \r) circle (2.5pt);
    \fill [black] (0,-\z) circle (2.5pt);
    \fill [green!60!black] (-\xx,-\yy) circle (2.5pt);
    \fill [green!60!black] (\xx,-\yy) circle (2.5pt);
 \end{tikzpicture}
 \;\;+\;\;
 \begin{tikzpicture}[valign,scale=0.8]
    \tikzstyle{every node}=[font=\scriptsize]
    \pgfmathsetmacro{\xx}{sqrt(2)/2}
    \pgfmathsetmacro{\yy}{sqrt(2)/2}
    \pgfmathsetmacro{\z}{0.5}
    \pgfmathsetmacro{\rx}{0.32}
    \pgfmathsetmacro{\ry}{0.54}
     \pgfmathsetmacro{\aa}{0.52}
      \pgfmathsetmacro{\bb}{-0.1}
    \draw [thick] (0,0) circle [radius=1];
    \draw      (-\xx,+\yy) -- (-\rx,-\ry);
    \draw      (+\xx,+\yy) -- (\rx,-\ry);
    \draw      (+\aa,-\bb) -- (-\aa,-\bb);
    \draw [thick, blue] (\xx,-\yy) to[out=150,in=30] (-\xx,-\yy);
     \fill [red!80!black] (-\xx,+\yy) circle (2.5pt);
     \fill [red!80!black] (+\xx,+\yy) circle (2.5pt);
    \fill [green!60!black] (-\rx,-\ry) circle (2.5pt);
    \fill [green!60!black] (\rx,-\ry) circle (2.5pt);
    \fill [black]  (+\aa,-\bb) circle (2.5pt);
    \fill [black] (-\aa,-\bb) circle (2.5pt);
 \end{tikzpicture}
 \;\;+\;\;
 \begin{tikzpicture}[valign,scale=0.8]
    \tikzstyle{every node}=[font=\scriptsize]
    \pgfmathsetmacro{\x}{-sqrt(2)/2}
    \pgfmathsetmacro{\y}{sqrt(2)/2}
    \pgfmathsetmacro{\ry}{0.1}
    \pgfmathsetmacro{\yy}{0.57}
    \pgfmathsetmacro{\xx}{0.45}
    \draw [thick] (0,0) circle [radius=1];
    \draw [thick, blue] (-\x,-\y) to[out=150,in=30] (+\x,-\y);
    \draw     (-\x,+\y) -- (-\xx,-\yy);
    \draw     (+\x,+\y) -- (\xx,-\yy);
       \fill [red!80!black]  (-\x,+\y) circle (2.5pt);
     \fill [red!80!black] (+\x,+\y) circle (2.5pt);
    \fill [black] (0,-\ry) circle (2.5pt);
    \fill [green!60!black] (-\xx,-\yy) circle (2.5pt);
    \fill [green!60!black] (\xx,-\yy) circle (2.5pt);
  \end{tikzpicture}
  \;\;+\;\;
  \begin{tikzpicture}[valign,scale=0.8]
    \tikzstyle{every node}=[font=\scriptsize]
    \pgfmathsetmacro{\x}{-sqrt(2)/2}
    \pgfmathsetmacro{\y}{sqrt(2)/2}
    \pgfmathsetmacro{\xx}{sqrt(2)/2}
    \pgfmathsetmacro{\yy}{sqrt(2)/2}
    \pgfmathsetmacro{\z}{0.5}
    \pgfmathsetmacro{\rx}{0.32}
    \pgfmathsetmacro{\ry}{0.54}
    \draw [thick] (0,0) circle [radius=1];
    \draw      (-\xx,+\yy) -- (-\rx,-\ry);
    \draw      (+\xx,+\yy) -- (\rx,-\ry);
    \draw  (-\rx,-\ry) to[out=100,in=80] (\rx,-\ry);
    \draw [thick, blue] (-\x,-\y) to[out=150,in=30] (+\x,-\y);
     \fill [red!80!black] (-\xx,+\yy) circle (2.5pt);
     \fill [red!80!black] (+\xx,+\yy) circle (2.5pt);
    \fill [green!60!black] (-\rx,-\ry) circle (2.5pt);
    \fill [green!60!black] (\rx,-\ry) circle (2.5pt);
 \end{tikzpicture}
    \;\;+\;\;
  \ldots\;.
\end{align}
Note that at this order there is also another piece $G^{(1)}_{\rm disc}$ which comes from the $1/c$ correction to the one-point function product $G^{(0)}_{\rm defect}$.\footnote{There is no correction to $G^{(0)}_{\rm bulk}$ because we have chosen the unit normalization for the defect-free two-point functions.} However, this piece is quite trivial because one-point functions are protected (they come from the $\frac{1}{2}$-BPS three-point function $\langle\mathcal{O}\mathcal{D}\mathcal{D} \rangle$) and can be computed in the free theory.

\vspace{0.5cm}

\noindent{\bf The correlator $\llangle \mathcal{O}_2\mathcal{O}_2\rrangle$.} In this paper, we focus on the one-loop corrections to the correlator which has the lowest KK levels $k_1=k_2=2$. Compared to correlators of higher KK-levels, this correlator has a few simplifying features. From (\ref{solscfWardid}), it follows that we can focus on the reduced correlator $\mathcal{H}_{22}$ which contains all the essential dynamical information. Moreover, due to the shifts in the quantum numbers in (\ref{reducedcorrelator}), the lowest weight reduced correlator is an R-symmetry singlet and therefore depends only on the conformal cross ratios $U$ and $V$. Equivalently, it is a symmetric function of the variables $(z,\zb)$, recall the definition \eqref{eq:cross_ratios}. We note that, when viewed as a four-point function, the correlator is invariant under exchanging the two identical giant graviton operators inserted at positions 3 and 4. At the level of the reduced defect correlator, this implies that $\H_{22}$ has the additional symmetry
\begin{align}\label{eq:H_crossing}
    \H_{22}(z,\zb)=\H_{22}(z',\zb')\,,
\end{align}
where $z'\equiv\frac{z}{z-1}$ and similarly for $\zb'$.

The reduced correlator admits a large-$c$ expansion analogous to \eqref{eq:G_large-c}, with the tree-level contribution at order $1/\sqrt{c}$ being given by \cite{Chen:2026ium}\footnote{
Note that since in \eqref{eq:G_large-c} we expand the correlator in $1/\sqrt{c}$ compared to $1/N$ as in \cite{Chen:2026ium}, several expressions in our work differ by factors of 2. In particular, we have $\H^{(1)}_{\rm here}=\frac12\H^{(1)}_{\rm there}$.
}
\begin{align}\label{eq:H1}
	\mathcal H^{(1)}_{22}
	=
	\frac{V\big(V^2-1-2V\log V\big)}
	     {U(1-V)^3}\,.
\end{align}
The next term, at order $1/c$, is the one-loop contribution $\H^{(2)}_{22}$ which is our main object of interest.

\vspace{0.5cm}

\noindent{\bf Superconformal block decomposition.}
The reduced correlator $\mathcal{H}_{22}$ admits an expansion into long superconformal blocks. The contribution from long multiplets takes a particularly simple form \cite{Chen:2026ium}: in the bulk channel, for a long superconformal multiplet whose super primary has dimension $\Delta$ and spin $\ell$, the contribution to the reduced correlator is simply the bosonic conformal block
\begin{equation}
    g_{\Delta+4,\ell}(z,\bar{z})\;,
\end{equation}
where 
\begin{equation}\label{eq:block_bulk}
  g_{\Delta,\ell}(z,\bar z)
  =
  \frac{(1-z)^2(1-\bar z)^2}{(z-\bar z)(z\bar z)^3}
  \left[
    k_{\Delta+\ell}(z) k_{\Delta-\ell-2}(\bar z)
    -
    k_{\Delta+\ell}(\bar z) k_{\Delta-\ell-2}(z)
  \right]\;,
\end{equation}
and
\begin{equation}
  k_\beta(z)=z^{\beta/2}\,
  {}_2F_1\!\left(\frac{\beta}{2},\frac{\beta}{2};\beta;z\right)\;.
\end{equation}
The shift of $\Delta\to\Delta+4$ is a consequence of superconformal symmetry which also shows up in the case of the all light four-point function.

In the defect channel, for a superconformal primary of a long multiplet with dimension $\widehat{\Delta}$ and transverse spin $s$, the contribution to $\mathcal{H}_{22}$ is simply
\begin{equation}
    \widehat{g}_{\widehat{\Delta}+2,s}(z,\bar{z})\;,
\end{equation}
where the defect channel conformal block is (assuming $0<V<1$)
\begin{equation}\label{eq:block_defect}
    \widehat{g}_{\widehat{\Delta},s}(z,\bar{z})= \big((1-z)(1-\zb)\big)^{\frac{\widehat{\Delta}-s}{2}}\,\frac{(1-\zb)^{s+1}-(1-z)^{s+1}}{z-\zb}\;,
\end{equation}
Although long multiplets contribute similarly to $\mathcal{H}_{k_1k_2}$ with higher values of $k_1,\,k_2$, the situation is complicated by the multiple R-symmetry channels which are absent in $\mathcal{H}_{22}$.

\vspace{0.5cm}

\noindent{\bf Mellin representation.} A convenient way to represent holographic correlators is the Mellin formalism \cite{Mack:2009mi,Penedones:2010ue}, where the analytic structure is simplified and the AdS scattering amplitude nature becomes more manifest. Here we will use its defect version \cite{Rastelli:2017ecj,Goncalves:2018fwx}. Note that, as was explained in \cite{Chen:2026ium}, the full giant graviton correlators do not admit a Mellin space representation. However, Mellin amplitudes can be defined for the reduced correlators. For the $k_1=k_2=2$ reduced correlator, the Mellin representation is defined as
\begin{equation}\label{eq:Mellin transform}
\mathcal{H}_{22}(z,\bar{z})
=
\int \frac{d\delta \,d\gamma}{(2\pi i)^2}
\, B^{-\delta}D^{\gamma+1}
\, \widetilde{\mathcal{M}}_{22}(\delta,\gamma)\,
\Gamma(\delta)\Gamma(\gamma-\delta+1)\Gamma\!\left(\frac{3-\gamma}{2}\right)^2 \,.
\end{equation}
In the above definition, we have also introduced the following combinations of cross ratios
\begin{equation}\label{eq:B_D_def}
B=\frac{U}{1+V-U}=\frac{z \bar z}{2-z-\bar z}\;,
\qquad
D=\frac{\sqrt{V}}{1+V-U}=\frac{\sqrt{(1-z)(1-\bar z)}}{2-z-\bar z} \;,
\end{equation}
which are the natural variables from the defect correlator perspective \cite{Chen:2024orp}. The bulk channel OPE corresponds to the limit $B\to 0$ and the defect channel OPE corresponds to $D\to0$.

For illustration, the Mellin amplitude of the tree-level correlator quoted in \eqref{eq:H1} is simply given by
\begin{align}\label{eq:Mellin-tree}
    \widetilde{\mathcal M}^{(1)}_{22}(\delta,\gamma) = \frac{1}{(\delta-1)(\gamma-1)}\,.
\end{align}
We will see later that the simultaneous pole structure is preserved also at one-loop level, although the number of poles becomes infinite.

\section{Unitarity method and leading logarithmic singularities}\label{section:unitarity_method}

\begin{figure}[t]
    \centering
    \includegraphics[width=\textwidth]{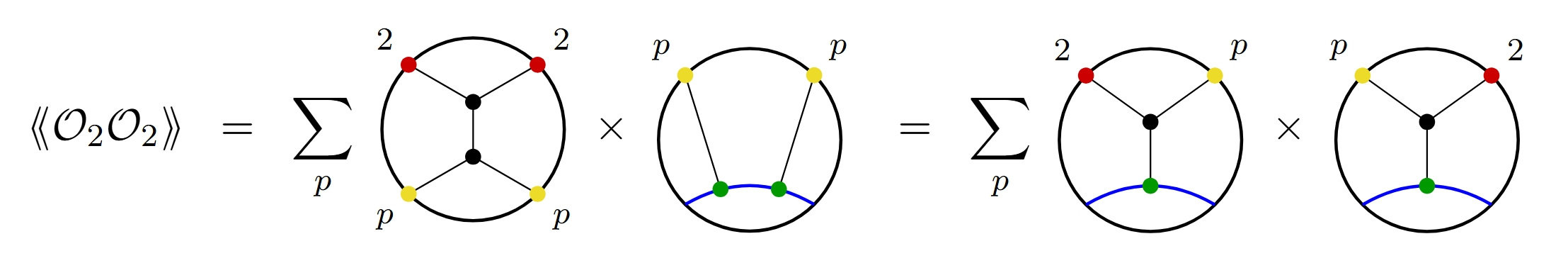}
    \caption{Schematic illustration of the AdS unitarity method for the one-loop defect two-point function. The defect correlator can be obtained by gluing defect-free tree-level four-point functions and disconnected defect two-point functions (bulk channel), or equivalently by gluing two copies of tree-level defect two-point functions (defect channel). In both cases one also needs to sum over all infinitely many KK modes which can run in the loop.}
    \label{fig:gluing_diagram}
\end{figure}

The one-loop correction to giant graviton correlators is difficult to compute directly. On the one hand, the direct computation involves complicated new Witten diagrams which would require significant improvement of the current techniques. On the other hand, all infinitely many Kaluza-Klein modes will appear in the loops and resumming their contributions appears to be a daunting task. Nevertheless, one can construct the one-loop correction via an analytic bootstrap approach which makes use of an AdS version of the flat-space unitarity method. Such a general AdS unitarity method was first developed in the defect-free context for four-point functions in \cite{Aharony:2016dwx}, and strongly coupled $\mathcal{N}=4$ SYM has been a main target of application, see, e.g.,  \cite{Alday:2017xua,Aprile:2017bgs,Aprile:2017qoy, Alday:2018kkw,Aprile:2019rep,Alday:2019nin,Huang:2021xws,Drummond:2022dxw}. Recently, this method is further extended to CFTs with conformal defects in \cite{Chen:2024orp} where defect two-point functions are considered in detail. The key idea of the method is schematically illustrated in Figure~\ref{fig:gluing_diagram}. The one-loop defect two-point function can be obtained by either gluing tree-level defect-free four-point functions with disconnected defect two-point functions, or gluing together tree-level defect two-point functions. 

\subsection{AdS unitarity method for defects}
More precisely, what gets `glued together' is the OPE data of correlators from lower orders in the large-$c$ expansion. The result of this procedure are certain leading logarithmic singularities (LLS) of the one-loop correlator, from which the full correlator can then be reconstructed.

To see how such logarithmic singularities arise, it is useful to consider an individual conformal block. In the bulk channel OPE limit, which corresponds to taking the small $U$ limit, a single bulk channel conformal block \eqref{eq:block_bulk} takes the form
\begin{equation}
    g_{\Delta,\ell}\big|_{U\to 0}\sim U^{\frac{\Delta-\ell}{2}}(1+\ldots)=U^{\frac{\Delta_0-\ell}{2}}\left(1+\frac{\gamma}{2}\log U+\ldots\right)\;,
\end{equation}
where in the second equality we have expanded the conformal dimension $\Delta=\Delta_0+\gamma$ with respect to a small anomalous dimension $\gamma$. Similarly, in the appropriate defect channel OPE limit for small $V$, the defect channel conformal block \eqref{eq:block_defect} behaves as
\begin{equation}
    \widehat{g}_{\widehat{\Delta},s}\big|_{V\to0}\sim V^{\frac{\widehat{\Delta}-s}{2}}(1+\ldots)= V^{\frac{\widehat{\Delta}_0-s}{2}}\left(1+\frac{\widehat{\gamma}}{2}\log V+\ldots\right)\;,
\end{equation}
where we have taken the defect conformal dimension $\widehat{\Delta}=\widehat{\Delta}_0+\widehat{\gamma}$ to have a small anomalous dimension $\widehat{\gamma}$.
In both cases, one can see that the logarithmic singularities which arise in the defect correlator are closely related to the anomalous dimensions of the exchanged unprotected operators.

For the giant graviton correlator in the supergravity limit, the relevant unprotected operators are double-particle operators. In the bulk channel, these operators take the schematic form ${:}\mathcal{O}_p \square^n\partial^\ell\mathcal{O}_p{:}$. They have conformal dimensions $2p+2n+\ell+\frac{1}{c}\gamma_{n,\ell} +\ldots$, where the leading anomalous dimensions are of order $\mathcal{O}(1/c)$. In the defect channel, the unprotected operators are double-particle operators formed out of a light supergraviton and a heavy giant graviton, where the heavy background is subtracted in the defect language \cite{Chen:2024orp}. They can be schematically denoted as $[\widehat{\mathcal{O}}_p]_{m,s}={:}\mathcal{O}_p\square^m\partial^s\mathcal{D}{:}/\mathcal{D}$, and have conformal dimensions $p+2m+s+ \frac{1}{\sqrt{c}} \widehat{\gamma}_{m,s}+\ldots$. It is important to note that the leading anomalous dimensions in the defect channel are of order $\mathcal{O}(c^{-1/2})$ rather than $\mathcal{O}(1/c)$. Therefore, in the large-$c$ expansion of the defect two-point correlators, we expect at most a single power of $\log V$ and no $\log U$ at tree level, whereas the one-loop correlator has at most a $\log^2 V$ and a $\log U$, see Figure~\ref{fig:logs}.
\begin{figure}[t]
\centering
\resizebox{0.62\textwidth}{!}{
\begin{tikzpicture}[cell/.style={draw=black!55,minimum width=3.0cm,minimum height=0.95cm,align=center,font=\small},note/.style={font=\small, align=center}]

\node[cell, fill=cyan!20]  at (0,0) {$\log U\,\log^2 V$};
\node[cell, fill=blue!12]  at (3.0,0) {$\log U\,\log V$};
\node[cell, fill=blue!12]  at (6.0,0) {$\log U$};

\node[cell, fill=green!12] at (0,-0.95) {$\log^2 V$};
\node[cell]  at (3.0,-0.95) {$\log V$};
\node[cell]  at (6.0,-0.95) {$1$};

\draw[blue!70!black, very thick, rounded corners=2pt](-1.50,0.475) rectangle (7.50,-0.475);

\draw[green!45!black, very thick, rounded corners=2pt](-1.50,0.475) rectangle (1.50,-1.425);

\node[draw=blue!70!black, thick, rounded corners=1pt, fill=blue!12, minimum width=0.35cm, minimum height=0.22cm, inner sep=0pt] at (-0.9,-2.05) {};
\node[note, blue!70!black, anchor=west] at (-0.6,-2.05){bulk-channel LLS};

\node[draw=green!45!black, thick, rounded corners=1pt, fill=green!12, minimum width=0.35cm, minimum height=0.22cm, inner sep=0pt] at (3.2,-2.05) {};
\node[note, green!45!black, anchor=west] at (3.5,-2.05){defect-channel LLS};

\end{tikzpicture}}
\caption{Structure of logarithmic terms of the one-loop correlator $\H^{(2)}$ as dictated by the bulk and defect channel OPE limits.}
\label{fig:logs}
\end{figure}

More explicitly, the defect channel LLS of the tree-level correlator reads
\begin{equation}
    \mathcal{H}^{(1)}\big|_{\log V}=\sum_{m,s} \frac{1}{2}\langle \widehat{A}^{(0)}_{m,s} \widehat{\gamma}^{(1)}_{m,s} \rangle \, \widehat{g}_{4+2m+s,s}\;.
\end{equation}
Here $\widehat{A}_{m,s}^{(0)} \equiv (\widehat b^{(0)}_{2,m,s})^2$ is defined from the  bulk-to-defect OPE coefficient $\widehat b_{2,m,s}^{(0)}$, and these are the defect channel OPE coefficients of the long double-particle multiplets in the free propagator term $\mathcal{G}^{(0)}_{\rm bulk}$. Note that the decomposition coefficients are in angle brackets, which denotes an implicit summation over degenerate operators. This is because in the infinite $c$ limit the double-particle operators described above are exactly degenerate and a given correlator only contains a particular average of their OPE data.

For the one-loop correlator, the LLS in the defect channel can be decomposed into conformal blocks as
\begin{equation}
    \mathcal{H}^{(2)}\big|_{\log^2 V}=\sum_{m,s} \frac{1}{8}\langle \widehat{A}^{(0)}_{m,s} (\widehat{\gamma}^{(1)}_{m,s})^2 \rangle \, \widehat{g}_{4+2m+s,s}\;.
    \label{eq:defectlog^2 V}
\end{equation}

On the other hand, in the bulk channel the LLS is given by
\begin{equation}
    \mathcal{H}^{(2)}\big|_{\log U}=\sum_{n,\ell} \frac{1}{2}\langle a^{(0)}_{n,\ell}\lambda^{(0)}_{2,n,\ell} \gamma^{(1)}_{n,\ell} \rangle \, g_{8+2n+\ell,\ell}\;.
\end{equation}
Here $a^{(0)}_{n,\ell}$ are the one-point function coefficients of the bulk channel double-particle operators which can be extracted from the disconnected piece $\mathcal{G}^{(0)}_{\rm defect}$. The coefficients $\lambda^{(0)}_{2,n,\ell}$ and $\gamma^{(1)}_{n,\ell}$ are encoded in the defect-free four-point functions 
\begin{align}\label{4ptdisc}
    \mathcal{H}_{\text{4pt},kkkk}^{(0)}\big|_{\rm long} &\supset \sum_{n,\ell} \langle (\lambda^{(0)}_{k,n,\ell})^2 \rangle \, G^{\text{4pt}}_{2k+2n+4+\ell,\ell}\;,\\
    \mathcal{H}_{\text{4pt},22kk}^{(1)}\big|_{\log U}&=\sum_{n,\ell} \frac{1}{2} \langle \lambda^{(0)}_{2,n,\ell}\lambda^{(0)}_{k,n,\ell}  \gamma^{(1)}_{n,\ell} \rangle \, G^{\text{4pt}}_{2k+2n+4+\ell,\ell}\;,\label{4ptlogU}
\end{align} 
where $\mathcal{H}_{\rm 4pt}^{(0)}$ and $\mathcal{H}_{\rm 4pt}^{(1)}$ are the reduced four-point correlators at disconnected and tree-level orders. $G^{\text{4pt}}_{\Delta,\ell}$ are the four-point function conformal blocks and are essentially the bulk channel defect conformal blocks  $g_{\Delta,\ell}$ up to an overall prefactor of cross ratios. Here we have also restricted to the singlet R-symmetry channel which is the only channel needed for computing the one-loop correction to $\llangle \mathcal{O}_2\mathcal{O}_2 \rrangle$. An important feature to notice is that all the data appearing in the decomposition coefficients of the one-loop LLS already appear at lower orders. If we temporarily ignore the complication of operator mixing, the one-loop LLS coefficients can be directly obtained from those at tree-level and disconnected orders, which makes the gluing construction plausible. As we will further show later in this section, the issue of operator mixing can be addressed by considering certain \textit{families} of correlators and therefore does not present an obstruction.

\vspace{0.5cm}
\noindent In summary, the strategy for computing the one-loop correction via the AdS unitarity method is a two-step procedure. First, we extract the lower-order OPE data to compute the LLS (for both the bulk and the defect channel). Second, we use these LLS as a constraint and complete them into the full one-loop correlator. The computation of the LLS is explicitly carried out in the next subsection, where the problem of operator mixing is also discussed. The completion of the LLS into the full one-loop correlator will then be done in later sections via two complementary approaches: in Section \ref{Sec:Mellinspace}, we compute the one-loop correlator in Mellin space and in Section \ref{Sec:positionspace} we perform the calculation in position space.

\subsection{Unmixing and resummation of leading logarithmic singularities}\label{Subsec:resumedLLS}

In this subsection, we discuss the problem of operator mixing and give explicit expressions for the LLS. The discussion is similar to the surface defect case in the 6d $(2,0)$ theory \cite{Chen:2024orp}. Some results have also been reported in \cite{Chen:2026ium}.

\vspace{0.5cm}

\noindent {\bf Defect channel unmixing.} Let us first consider the mixing problem in the defect channel. The double-particle operators 
\begin{equation}
\big\{[\widehat{\mathcal O}_2]_{L-1,s},\,[\widehat{\mathcal O}_4]_{L-2,s},\,\ldots,\,[\widehat{\mathcal O}_{2L}]_{0,s}\big\}\;,\qquad L=1,2,\ldots\;,
\end{equation}
all have the same degenerate defect twist $2L$ at infinite $c$. Recall that the operators in this degenerate family are of the form $[\widehat{\mathcal O}_p]_{m,s}$ with $p+2m=2L$, so the parity of $p$ is fixed at fixed twist. We are interested in the correlator $\llangle\mathcal O_2\mathcal O_2\rrangle$, so that only $p=2,4,\ldots,2L$ can appear here.

It is convenient to organize the bulk-to-defect OPE coefficients into a matrix. Let us use the notation $\{\widehat O_1,\ldots,\widehat O_{L}\}$ for a new basis of these operators where the defect channel tree-level anomalous dimensions are diagonalized. We then define
\begin{equation}
\widehat{\bf B}=\left(
\begin{array}{cccc}
\widehat b^{(0)}_{2\widehat O_1} &\widehat b^{(0)}_{2\widehat O_2} &\ldots &\widehat b^{(0)}_{2\widehat O_{L}}\\
\widehat b^{(0)}_{4\widehat O_1} &\widehat b^{(0)}_{4\widehat O_2} &\ldots &\widehat b^{(0)}_{4\widehat O_{L}}\\
\ldots & \ldots & \ldots & \ldots\\\widehat b^{(0)}_{2L\widehat O_1} &\widehat b^{(0)}_{2L\widehat O_2} &\ldots &\widehat b^{(0)}_{2L\widehat O_{L}}
\end{array}\right) \;,
\end{equation}
where $k$ in the OPE coefficient $\widehat b^{(0)}_{k\widehat O_m}$ labels the external operator $\mathcal{O}_k$ inserted in the bulk. Note that these defect double-particle operators are R-symmetry singlets, and this requires $k$ to be even. In terms of this matrix, we can conveniently write down the defect channel decomposition coefficients of the zeroth-order correlator $G^{(0)}_{\rm bulk}$ coming from the free bulk propagator. These coefficients correspond to a diagonal matrix
\begin{equation}\label{BBTeqN}
\widehat{\bf B}\widehat{\bf B}^{T}=\widehat{\bf N}={\rm diag}\{\llangle \mathcal O_2\mathcal O_2\rrangle^{(0)}_{\rm bulk},\llangle \mathcal O_4\mathcal O_4\rrangle^{(0)}_{\rm bulk},\ldots,\llangle \mathcal O_{2L}\mathcal O_{2L}\rrangle^{(0)}_{\rm bulk}\}\; .
\end{equation}
Let us also define another diagonal matrix of which the elements are the tree-level defect anomalous dimensions
\begin{equation}
\widehat{\bf \Gamma}={\rm diag}\{\widehat{\gamma}^{(1)}_1,\ldots,\widehat{\gamma}^{(1)}_{L}\}\;.
\end{equation}
The coefficient of the tree-level LLS (the $\log V$ term) then corresponds to
\begin{equation}\label{OmegaeqBGBT}
\widehat{\bf \Omega}=\widehat{\bf B}\widehat{\bf \Gamma}\widehat{\bf B}^{T}=\left(
\begin{array}{ccc}
\llangle \mathcal O_2\mathcal O_2\rrangle &
\ldots &
\llangle \mathcal O_2\mathcal O_{2L}\rrangle
\\
\ldots & \ldots & \ldots
\\
\llangle \mathcal O_{2L}\mathcal O_2\rrangle &
\ldots &
\llangle \mathcal O_{2L}\mathcal O_{2L}\rrangle
\end{array}
\right)^{\rm tree}_{\log V}\;.
\end{equation}
Similarly, the defect channel LLS at one loop ($\log^2 V$ term) can be organized into a similar matrix and can be expressed as
\begin{equation}\label{1loopmatrix}
\widehat{\bf B}(\widehat{\bf \Gamma})^2\widehat{\bf B}^{T}=\widehat{\bf \Omega}\,\widehat{\bf N}^{-1}\widehat{\bf \Omega}\;,
\end{equation}
where we have used (\ref{BBTeqN}) and (\ref{OmegaeqBGBT}) to rewrite it in terms of the tree-level and GFF matrices.

One could just proceed by diagonalizing these matrices and thereby solving the mixing problem. This has been carried out in \cite{Chen:2026ium} where the spectrum of anomalous dimensions was also obtained. Here our focus is the $\log^2V$ coefficient of the one-loop $\llangle \mathcal O_2\mathcal O_2\rrangle$. This can be read off from the top-left component of the correlator matrix (\ref{1loopmatrix}) without explicitly solving the mixing problem. Written explicitly in terms of matrix elements, we have
\begin{equation}
\frac{1}{8}
\langle
\widehat{A}^{(0)}_{m,s}
(\widehat{\gamma}^{(1)}_{m,s})^2
\rangle
=
\sum_{k=2,\rm{even}}^{2L}
\frac{
\langle
\widehat b^{(0)}_{2,m_{\rm d},s}
\widehat b^{(0)}_{k,m_{\rm d},s}
\widehat{\gamma}^{(1)}_{m_{\rm d},s}
\rangle^2
}{
8\,
\langle
( \widehat b^{(0)}_{k,m_{\rm d},s})^2
\rangle
}\;,
\qquad
m_{\rm d}=\frac{2L-k}{2} \;,
\end{equation}
where we have used the fact that the matrix $\widehat{\bf N}$ is diagonal. Now $L$ is identified as $m+1$. This expression makes it clear that the defect channel one-loop LLS can be computed entirely from lower order data. The long part of $G^{(0)}_{\text{bulk},kk}$ is decomposed as
\begin{equation}
G^{(0)}_{\text{bulk},kk}\big|_{\rm long}\supset R(V\sigma \tau)^{-1} \sum_{m,s=0}^{\infty}\langle (\widehat b^{(0)}_{k,m,s})^2\rangle \widehat{g}_{k+2m+2+s,s}\;,
\end{equation}
with coefficients 
\begin{equation}
    \langle (\widehat b^{(0)}_{k,m,s})^2\rangle  =  \frac{16(s+1)\,(m+1)_k\,(m+s+2)_k}{(k+2m)(k+2+2m)\big(k+2(1+m+s)\big)\big(k+2(2+m+s)\big)\Gamma\!\left(\frac{k}{2}+1\right)^2\Gamma\!\left(\frac{k}{2}\right)^2}\;.
\end{equation}
The decomposition of the  tree-level $\log V$ coefficients of
$\llangle \mathcal O_2\mathcal O_k\rrangle_{\rm tree}$ reads 
\begin{equation}
    \mathcal{H}^{(1)}_{2k} \big|_{\log V} = \sum_{m,s=0}^{\infty}\frac{1}{2}   
\langle
 \widehat b^{(0)}_{2,m,s}
 \widehat b^{(0)}_{k,m,s}
\widehat{\gamma}^{(1)}_{m,s}
\rangle\,\widehat{g}_{k+2m+2+s,s}\;,
\end{equation}
where 
\begin{equation}
    \langle
 \widehat b^{(0)}_{2,m,s}
 \widehat b^{(0)}_{k,m,s}
\widehat{\gamma}^{(1)}_{m,s}
\rangle = -\frac{\sqrt{2k}(m+1)_k}{\Gamma(\frac{k}{2}+1)\Gamma(\frac{k}{2})}\;.
\end{equation}
These results have already been reported in \cite{Chen:2026ium}, and the one-loop $\log^2 V$ coefficients are
\begin{equation}\label{defectunitarity}
    \frac{1}{8}
\langle
\widehat{A}^{(0)}_{m,s}
(\widehat{\gamma}^{(1)}_{m,s})^2
\rangle =  \sum_{k=2,\,\rm{even}}^{2m+2} \frac{
k(m+1)_2(m+s+2)_2(m-\frac{k}{2}+2)_{1+s}
}{
4(s+1)(m+\frac{k}{2}+2)_{1+s}
}\;.
\end{equation}

Before presenting the resummed expressions for the conformal block decomposition of the LLS, let us explain the choice of variables we will use in the two OPE channels. The Bulk OPE limit corresponds to $U\to0$, we therefore expand in small $z$, while keeping the full dependence on $\bar z$ in the bulk-channel resummation. For the defect channel, it is then useful to pass to the variables $x=1-z,\bar x=1-\bar z$. The defect OPE limit is $V\to0$, and the defect-channel resummation is naturally organized as a small-$x$ expansion, keeping the full dependence on $\bar x$. For convenience, we recall that the one-loop LLS in the defect channel is given by the infinite sum over defect channel conformal blocks
\begin{equation}
\mathcal{H}^{(2)}_{22}\big|_{\log^2 V}=\sum_{m,s=0}^{\infty}\frac{1}{8}\langle\widehat A^{(0)}_{m,s}\big(\widehat\gamma^{(1)}_{m,s}\big)^2\rangle\,\widehat g_{4+2m+s,s}(x,\bar x)\;.\label{eq:defect_LLS_sum}
\end{equation}
After substituting \eqref{defectunitarity}, our goal is to perform the double sum over the defect twist label $m$ and transverse spin $s$, and rewrite it as a closed-form function of $x$, $\bar{x}$.

\vspace{0.4cm}

 \noindent {\bf Defect channel LLS.} To perform the resummation in (\ref{eq:defect_LLS_sum}), we proceed by first expanding in small $x$. We keep the full $\bar{x}$ dependence at each fixed order in $x$, and analyze the pattern of these coefficients as the order increases. We are led to the following closed-form result for the leading-logarithmic defect-channel correlator
\begin{equation}\label{eq:defect resum}
\mathcal{H}^{(2)}_{22}(x,\bar x)\big|_{\log^2 V}
=\frac{x^2\bar x^2}{(x-\bar x)^7(1-x\bar x)^7}
\Big[P^{\rm def}_{0}(x,\bar x)-P^{\rm def}_{1}(x,\bar x)
\log\!\left(\frac{1-x}{1-\bar x}\right)\Big]\;.
\end{equation}
Here \(P^{\rm def}_{0}(x,\bar x)\) is antisymmetric under
\(x\leftrightarrow \bar x\), while \(P^{\rm def}_{1}(x,\bar x)\)
is symmetric. Both these polynomials are of bounded degrees. The explicit
expressions for these polynomials are collected in
Appendix~\ref{appendix:resum poly}.
This bounded polynomial structure is highly nontrivial from the viewpoint of the original expansion in terms of conformal blocks and provides a strong indication that the resummed correlator admits a compact analytic form. We have checked \eqref{eq:defect resum} by expanding it at small $x$ and comparing with \eqref{eq:defect_LLS_sum} to high order, providing strong evidence for its exactness.

\vspace{0.4cm}

\noindent {\bf Bulk channel unmixing.}
We now turn to the unmixing analysis in the bulk channel. The relevant exchanged operators are the double-particle operators
\begin{equation}
\big\{
:\mathcal O_2\square^{M-2}\partial^\ell\mathcal O_2:,\,
:\mathcal O_3\square^{M-3}\partial^\ell\mathcal O_3:,\,
\ldots,\,
:\mathcal O_M\partial^\ell\mathcal O_M:
\big\}\;,\qquad M=2,3,\ldots\; ,
\end{equation}
where we have fixed the twist at infinite $c$ to be $2M$. We also define the basis $\{O_1,\ldots,O_{M-1}\}$ which diagonalizes the anomalous dimensions. We then organize the leading order three-point coefficients into the matrix
\begin{equation}
{\bf \Lambda}=
\left(
\begin{array}{cccc}
\lambda^{(0)}_{22O_1} &
\lambda^{(0)}_{22O_2} &
\ldots &
\lambda^{(0)}_{22O_{M-1}}
\\
\lambda^{(0)}_{33O_1} &
\lambda^{(0)}_{33O_2} &
\ldots &
\lambda^{(0)}_{33O_{M-1}}
\\
\ldots & \ldots & \ldots & \ldots
\\
\lambda^{(0)}_{MMO_1} &
\lambda^{(0)}_{MMO_2} &
\ldots &
\lambda^{(0)}_{MMO_{M-1}}
\end{array}
\right)\;.
\end{equation}
The long operator contributions in disconnected defect-free four-point functions \eqref{4ptdisc} correspond to a diagonal matrix and can be expressed in terms of OPE coefficient matrix as 
\begin{equation}\label{LLTeqN}
{\bf \Lambda}{\bf \Lambda}^{T}
=
{\bf N}
=
{\rm diag}\{
\langle \mathcal O_2\mathcal O_2\mathcal O_2\mathcal O_2\rangle,\ldots,
\langle \mathcal O_M\mathcal O_M\mathcal O_M\mathcal O_M\rangle
\}_{\rm disc}\;.
\end{equation}
Let us also define the diagonal matrix of the tree-level anomalous dimensions of $O_i$
\begin{equation}
    {\bf \Gamma}={\rm diag}\{\gamma^{(1)}_1,\ldots,\gamma^{(1)}_{M-1}\}\;,
\end{equation}
and denote their one-point function coefficients as the vector
\begin{equation}
{\bf A}=(a^{(0)}_1,\ldots,a^{(0)}_{M-1})^T \;.
\end{equation}
 Then the disconnected part of the defect two-point function coming from the product of one-point functions have decomposition coefficients which correspond to the vector 
 \begin{equation}\label{coeffLA}
{\bf \Lambda}{\bf A}=\big(\llangle \mathcal O_2\mathcal O_2\rrangle,\ldots,\llangle \mathcal O_M\mathcal O_M\rrangle\big)^T_{G^{(0)}_{\rm defect}}\;,
\end{equation}
and the tree-level $\log U$ coefficient of the defect-free four-point function $\langle\mathcal{O}_p\mathcal{O}_p \mathcal{O}_q \mathcal{O}_q\rangle$ can be packaged into the matrix
\begin{equation}\label{OmegaeqLGLT}
\nonumber
{\bf \Omega}
=
{\bf \Lambda}{\bf \Gamma}{\bf \Lambda}^{T}
=
\left(
\begin{array}{ccc}
\langle \mathcal O_2\mathcal O_2\mathcal O_2\mathcal O_2\rangle &
\ldots &
\langle \mathcal O_2\mathcal O_2\mathcal O_M\mathcal O_M\rangle
\\
\ldots & \ldots & \ldots
\\
\langle \mathcal O_M\mathcal O_M\mathcal O_2\mathcal O_2\rangle &
\ldots &
\langle \mathcal O_M\mathcal O_M\mathcal O_M\mathcal O_M\rangle
\end{array}
\right)^{\rm tree}_{\log U}\;.
\end{equation}
 At one-loop level, the LLS coefficient of the defect two-point function can be expressed in the same way by inserting in (\ref{coeffLA}) another anomalous dimension matrix
\begin{equation}
{\bf \Lambda}{\bf \Gamma}{\bf A}={\bf \Omega}\,{\bf N}^{-1}({\bf \Lambda}{\bf A}) \;.
\end{equation}
On the RHS, we have rewritten the expression by using (\ref{LLTeqN}) and (\ref{OmegaeqLGLT}) to manifest the fact that it only depends on lower order data. Focusing on the $\llangle \mathcal O_2\mathcal O_2\rrangle$ correlator, which corresponds to the first component of the vector, we can explicitly write the conformal block decomposition coefficients as
\begin{equation}
\begin{aligned}
\label{bulkunitarity}
\frac{1}{2}\langle a^{(0)}_{n,\ell}\lambda^{(0)}_{2,n,\ell}\gamma^{(1)}_{n,\ell}\rangle
&=\mathop{\sum_{k=2}}_{k\,{\rm even}}^{M}\frac{\langle a^{(0)}_{m_{\rm b},\ell} \lambda^{(0)}_{k,m_{\rm b},\ell}\rangle\langle\lambda^{(0)}_{2,m_{\rm b},\ell}\lambda^{(0)}_{k,m_{\rm b},\ell}\gamma^{(1)}_{m_{\rm b},\ell}\rangle}{2\langle(\lambda^{(0)}_{k,m_{\rm b},\ell})^2\rangle}\;,
\end{aligned}
\end{equation}
with
\begin{equation}
m_{\rm b}=M-k = n+2-k\; .
\end{equation}
The defect-free part of the lower order coefficients are explicitly given by \cite{Aprile:2017xsp,Aprile:2019rep}
\begin{equation}
 \begin{aligned}
(\lambda^{(0)}_{k,m_{\rm b},\ell})^2  = {}&\frac{
24(1+\ell)(6+\ell+2n)\Gamma(n+1)\,
\Gamma(3+n)^2
\Gamma(3+k+n)
\Gamma(2+\ell+n)
}{
k^2(1+k)\Gamma(-1+k)\Gamma(k)^3
\Gamma(5+n)
\Gamma(3-k+n)
}
\\
&\times
\frac{
\Gamma(4+\ell+n)^2
\Gamma(4+k+\ell+n)
}{
\Gamma(6+\ell+n)
\Gamma(4-k+\ell+n)
\Gamma(5+2n)
\Gamma(7+2\ell+2n)
}\; ,
\end{aligned}
\end{equation}
\begin{equation}
  \langle
\lambda^{(0)}_{2,m_{\rm b},\ell}
\lambda^{(0)}_{k,m_{\rm b},\ell}
\gamma^{(1)}_{m_{\rm b},\ell} 
\rangle   =   -\frac{
2(-1)^k
\Gamma(3+n)^2
\Gamma(4+\ell+n)^2
(3-k+n)_{2k}
}{
\Gamma(-1+k)\Gamma(k)
\Gamma(5+2n)
\Gamma(7+2\ell+2n)
}\;.
\end{equation}
By focusing on the long part of the disconnected correlator $\llangle \mathcal{O}_k\mathcal{O}_k\rrangle^{(0)}_{\rm defect}$,
\begin{equation}
    G^{(0)}_{\text{defect},kk}\big|_{\rm long} \supset   R(V\sigma \tau)^{-1} \sum_{m=-k,\,\ell\, \rm{even}}^{\infty} \langle
a^{(0)}_{m,\ell}
\lambda^{(0)}_{k,m,\ell}
\rangle\, g_{2k+2m+4+\ell,\ell}\;,
\end{equation}
we can also extract the decomposition coefficients 
\begin{equation}
  \begin{aligned}
\langle
a^{(0)}_{m,\ell}
\lambda^{(0)}_{k,m,\ell}
\rangle = {}&\frac{
12(-1)^m 2^{-5-2k-2\ell-4m}\pi\,
\Gamma(1+k+m)\Gamma(2+k+\ell+m)
}{
k(1+k)\Gamma(k)^2
\Gamma(\frac12+k+m)
\Gamma(\frac32+k+\ell+m)
}
\\
&\times
(1+\lfloor \tfrac{m}{2}\rfloor)_{\frac{k-2}{2}}
(\tfrac{5+k+\ell}{2}+\lfloor \tfrac{m}{2}\rfloor)_{\frac{k-2}{2}}
\\
&\times
(\tfrac{3+k}{2}+\lfloor \tfrac{1+m}{2}\rfloor)_{\frac{k-2}{2}}
(1+\tfrac{\ell}{2}+\lfloor \tfrac{1+m}{2}\rfloor)_{\frac{k-2}{2}} \;.
\end{aligned}
\end{equation}
Here we recall that $k$ must be even for the one-point function, and therefore $\llangle \mathcal{O}_k\mathcal{O}_k\rrangle^{(0)}_{\rm defect}$, to be non-vanishing, which explains why the sum in \eqref{bulkunitarity} runs only over even values of $k$. Also note that only the R-symmetry singlet channel is needed. 

The remaining task is to perform the infinite sum over bulk-channel conformal blocks after substituting \eqref{bulkunitarity} into
\begin{equation}
\mathcal{H}^{(2)}_{22}\big|_{\log U}=\sum_{n,\ell=0}^{\infty}\frac{1}{2}\langle a^{(0)}_{n,\ell}\lambda^{(0)}_{2,n,\ell}\gamma^{(1)}_{n,\ell}\rangle\, g_{8+2n+\ell,\ell}(z,\bar z)\;.
\label{eq:bulk_LLS_sum}
\end{equation}

\vspace{0.4cm}

\noindent {\bf Bulk channel LLS.} The resummation of LLS in the bulk channel can be analogously done as in the defect channel. We first expand in small $z$ near the bulk channel OPE limit, keeping the full dependence on $\bar{z}$. After identifying the transcendental weight pattern from the low-order expansions, we arrive at the following general form of the bulk channel LLS
\begin{equation}
\begin{split}
\mathcal{H}^{(2)}_{22}(z,\bar z)\big|_{\log U}
&=\frac{(1-z)^2(1-\bar z)^2}{(z-\bar z)^7(z+\bar z-z\bar z)^7}
\Bigg[
P^{\rm bulk}_{0}(z,\bar z)
+P^{\rm bulk}_{1}(z,\bar z)\log(1-z)\\
&\quad-P^{\rm bulk}_{1}(\bar z,z)\log(1-\bar z)
-2P^{\rm bulk}_{2}(z,\bar z)
\log\!\left(\frac{1-z}{1-\bar z}\right)
\log\!\big((1-z)(1-\bar z)\big)
\Bigg]\;,
\end{split}
\label{eq:bulk resum}
\end{equation}
where  $P^{\rm bulk}_i(z,\bar z)$ are various polynomials in $z$ and $\bar{z}$. In particular, $P^{\rm bulk}_{0}(z,\bar z)$ is antisymmetric under $z\leftrightarrow\bar z$ and $P^{\rm bulk}_{2}(z,\bar z)$ is symmetric. The explicit expressions of these polynomials are given in Appendix~\ref{appendix:resum poly} and we checked the closed form expression to high orders in expansions. Moreover, a very nontrivial consistency check of the LLS is that the coefficient of $\log U\log^2V$ is the same in both (\ref{eq:defect resum}) and (\ref{eq:bulk resum}).

\section{The one-loop correlator in Mellin space}\label{Sec:Mellinspace}

In this section, we compute the full one-loop reduced correlator in Mellin space, using as input the leading logarithmic singularities from Section \ref{section:unitarity_method}. We use the variables $B,D$ introduced in \eqref{eq:B_D_def}, which are adapted to the defect Mellin representation. Before diving into the details of the calculation, let us first outline the general strategy which consists of the following steps:  
\begin{enumerate}
\item We start by making an ansatz for the reduced Mellin amplitude which features simultaneous simple poles\footnote{\label{foot:tree-level_amb}Here we restrict ourselves to adding only poles above the two-particle threshold, i.e. poles which overlap with the ones already present in the gamma functions in the definition of the Mellin amplitude \eqref{eq:Mellin transform}. The only possible additional (simple) poles one can add are at $\delta=\gamma=1$. This corresponds precisely to the tree-level amplitude \eqref{eq:Mellin-tree}, which does not contribute to the LLS at one-loop order and therefore can be added with arbitrary coefficient to the one-loop amplitude. We will come back to this ambiguity later on in Section \ref{Sec:positionspace}.}
\begin{equation}
\widetilde{\mathcal{M}}^{(2)}_{22}(\delta, \gamma) = \sum_{m,n=0}^{\infty} \frac{c_{mn}}{(\delta + n)(\gamma - 3 - 2m)} \,. \label{eq:MellinAnsatz}
\end{equation}
Such simultaneous poles are needed in order to reproduce the simultaneous leading logarithmic singularities $\log B\,\log^2D$. Notice that the gamma function factor already contains poles at these locations. Therefore, first order poles in the reduced Mellin amplitude are sufficient. 
\item Then we make a seemingly radical assumption that the $c_{mn}$ coefficients in the numerators are just constants. In other words, we assume that the reduced Mellin amplitude does not contain any single poles. 
\item We can determine the coefficients $c_{mn}$ by comparing the simultaneous LLS computed in two ways:\footnote{
When switching from $(U,V)$ variables to $(B,D)$ via \eqref{eq:B_D_def}, the logarithmic structure is affected as follows. For the bulk channel OPE, we have that $\log B\sim\log U+ (\text{regular terms})$, so the leading-log coefficient can be compared directly between Mellin space and position space. For the defect OPE, one has $\log D\sim\frac12\log V+ (\text{regular terms})$, which gives a factor of 4 when comparing the $\log^2D$ coefficient in Mellin space with the $\log^2V$ coefficient in position space. The regular terms only affect lower-logarithmic contributions and do not enter our determination of the coefficients $c_{mn}$, although they must be kept when comparing subleading logarithmic terms.}
from taking residues in Mellin space with the ansatz (\ref{eq:MellinAnsatz}), and from the position space resummations obtained in Section \ref{Subsec:resumedLLS}. In Figure \ref{fig:log-power-table}, this corresponds to matching the coefficient function of $\log B\,\log^2D$.
\item After obtaining $c_{mn}$ in a closed form, we perform a partial resummation (either in $m$ or in $n$) and compute the full LLS  in both the bulk and the defect channel. Comparing with the position space result, we find that the subleading logarithmic terms $\log^2D$, $\log B\,\log D$, and $\log B$ in Figure \ref{fig:log-power-table} are also matched. This justifies our assumption that additional single poles are absent.
\item As is clear from Figure \ref{fig:log-power-table}, we have now fixed the entire singular part of the reduced Mellin amplitude. The remaining ambiguity can only be regular in the Mellin variables, and the flat-space limit constrains this to be just a constant which has a natural interpretation as a one-loop counter-term ambiguity.
\end{enumerate}
\begin{figure}[t]
\centering
\resizebox{0.64\textwidth}{!}{
\begin{tikzpicture}[
    cell/.style={draw=black!55, minimum width=3.0cm, minimum height=0.95cm,
      align=center, font=\small},
    note/.style={font=\small, align=left}
]
% table body
\node[cell, fill=red!12]  (a11) at (0,0)
  {$\log B\,\log^2D$};
\node[cell, fill=blue!12] (a12) at (3.0,0)
  {$\log B\,\log D$};
\node[cell, fill=blue!12] (a13) at (6.0,0)
  {$\log B$};

\node[cell, fill=green!12] (a21) at (0,-0.95)
  {$\log^2D$};
\node[cell, fill=gray!10]  (a22) at (3.0,-0.95)
  {$\log D$};
\node[cell, fill=gray!10]  (a23) at (6.0,-0.95)
  {$1$};

% frames
\draw[blue!70!black, very thick, rounded corners=2pt]
  (-1.50,0.475) rectangle (7.50,-0.475);

\draw[green!45!black, very thick, rounded corners=2pt]
  (-1.50,0.475) rectangle (1.50,-1.425);

\draw[red!75!black, very thick, rounded corners=2pt]
  (-1.42,0.395) rectangle (1.42,-0.395);

\node[draw=blue!70!black, thick, rounded corners=1pt, fill=blue!12, minimum width=0.35cm, minimum height=0.22cm, inner sep=0pt] at (-1.3,-2.0) {};
\node[note, blue!70!black, anchor=west] at (-1.0,-2.0){bulk-channel LLS data};

\node[draw=green!45!black, thick, rounded corners=1pt, fill=green!12, minimum width=0.35cm, minimum height=0.22cm, inner sep=0pt] at (3.4,-2.0) {};
\node[note, green!45!black, anchor=west] at (3.7,-2.0){defect-channel LLS data};

\node[draw=red!75!black, thick, rounded corners=1pt, fill=red!12, minimum width=0.35cm,
  minimum height=0.22cm, inner sep=0pt, anchor=west] at (-1.5,-2.6) {};
\node[note, red!75!black, anchor=west] at (-1.0,-2.6)
  {terms used to determine the Mellin coefficients $c_{mn}$};

\node[draw=black!55, thick, rounded corners=1pt, fill=gray!10, minimum width=0.35cm,
  minimum height=0.22cm, inner sep=0pt, anchor=west] at (-1.5,-3.2) {};
\node[note, black!70, anchor=west] at (-1.0,-3.2)
  {remaining lower-log terms fixed by the Mellin residues};

\end{tikzpicture}
}
\caption{Structure of logarithmic terms of the one-loop correlator in $(B,D)$ variables.}
\label{fig:log-power-table}
\end{figure}

\subsection{Determining the Mellin coefficients $c_{mn}$}

We start by solving for the $c_{mn}$ coefficients in the ansatz \eqref{eq:MellinAnsatz}. As we explained in the outline of the algorithm, these coefficients are fixed by comparing the coefficient function of $\log^2D\,\log B$ computed from Mellin space with those from the position space resummation. In practice, we can proceed in two ways which give difference slices of the $c_{mn}$ coefficients. 

In the first route, we analyze the Mellin space integral in the bulk OPE limit where $B\to 0$. We take the residue at $\delta=-n$ and focus on the $\log B$ term. The residues are distinguished by different powers $B^n$. We then take the residue at $\gamma=3+2m$ and isolate the $\log^2 D$ coefficient, where the residue at each pole is multiplied by $D^{4+2m}$. We expand the position space result of the $\log^2D\,\log B$ coefficient in the same way and require the coefficients to match for each power $B^nD^{4+2m}$. For fixed values of $n$, it is easy to find the general functional dependence on $m$. The first few families of coefficients are 
\begin{align}
\begin{split}
c_{m0}&=-\frac{(6m+8)(1)_m}{12\left(\frac52\right)_m}\;,\\
c_{m1}&=-\frac{(5m+6)\,(1)_m}{12\,\left(\frac52\right)_m}\;,\\
c_{m2}&=-\frac{(25m^2+79m+48)\,(1)_m}{120\,\left(\frac72\right)_m}\;,\\
c_{m3}&=-\frac{(57m^2+161m+80)\,(1)_m}{240\,\left(\frac72\right)_m}\;.
\end{split}
\end{align}
The obvious pattern in the coefficients continue to hold for higher values of $n$ and the degree of the polynomial in the numerator grows as $\lfloor\frac{n+1}{2}\rfloor$.

In the second route, we interchange the order of expansion in $B$ and $D$. This implies that $m$ is fixed but $n$ is arbitrary. We can also find general expressions for the coefficients in this slicing. For low lying values of $m$, the coefficients are 
\begin{align}\label{eq:Low-lying defect-channel data}
\begin{split}
c_{0n}
&=
-\frac{2}{n+3} \;,
\\
c_{1n}
&=
-\frac{1}{n+1}
+\frac{4}{n+2}
-\frac{8}{n+3}
+\frac{24}{n+4}
-\frac{24}{n+5} \;,
\\
c_{2n}
&=
-\frac{2}{n+1}
+\frac{20}{n+2}
-\frac{74}{n+3}
+\frac{168}{n+4}
-\frac{300}{n+5}
+\frac{360}{n+6}
-\frac{180}{n+7} \;,
\\
c_{3n}
&=
-\frac{3}{n+1}
+\frac{48}{n+2}
-\frac{290}{n+3}
+\frac{972}{n+4}
-\frac{2178}{n+5}
+\frac{3640}{n+6}
-\frac{4440}{n+7}
+\frac{3360}{n+8}
-\frac{1120}{n+9} \;.
\end{split}
\end{align}
These coefficients exhibit a clear pattern as a sum over poles in $n$ where the number of poles  increases as $2m+3$. 

Computing more of these coefficients allows us to identify further patterns in them. Through trial and error, we find that the general $c_{mn}$ coefficient admits the following closed form expression:
\begin{equation}\label{eq:generalCmn}
c_{mn}=\frac{2^{2m-5}\sqrt{\pi}\,\Gamma(n+m-1)}{3\,\Gamma\!\left(\frac52-m\right)\Gamma(n+2m-2)}\Big[\alpha_{m,n}\, \mathfrak h_{-1}(m,n)+\beta_m\, \mathfrak h_{2}(m,n)+\gamma_{m,n}\, \mathfrak h_{0}(m,n)\Big]\,.
\end{equation}
The functions $\mathfrak h_a(m,n)$ are a class of hypergeometric building blocks defined by
\begin{equation}
\mathfrak h_a(m,n)={}_3F_2\!\left(\begin{matrix}
a-m\,,\; a-m-\frac{n+1}{2}\,,\; a-m-\frac n2\,\\
\frac12+a-m\,,\;a-m-n
\end{matrix}\Bigg|\,1\right)\,.
\end{equation}
Their coefficient functions are given by
\begin{align}
\begin{split}
\alpha_{m,n}&=-\frac{2(2m-3)(2m-1)(2m+1)(n+m-1)_3}{\prod_{k=-2}^{3}(2m+n+k)}\,\mathsf p_1(m,n)\;,\\[5pt]
\beta_m&=20(m-1)^2m^2 \;,\\[5pt]
\gamma_{m,n}&=-\frac{(2m-3)(2m-1)(n+m-1)_2}{\prod_{k=-2}^{1}(2m+n+k)}\,\mathsf p_2(m,n) \;,
\end{split}
\end{align}
where the polynomials $\mathsf p_1(m,n)$ and $\mathsf p_2(m,n)$ take the form
\begin{align}
\begin{split}
\mathsf p_1(m,n)&=160m^4+40m^3(6n+7)+4m^2(35n^2+65n-17)\\
&\quad+4m(10n^3+25n^2-22n-47)+n(5n^3+20n^2-3n-46)\;,
\end{split}\\[3pt]
\begin{split}
\mathsf p_2(m,n)&=240m^4+120m^3(2n+1)+4m^2(35n^2+35n-27)\\
&\quad+4m(10n^3+15n^2-17n-26)+(5n^4+10n^3-3n^2-20n+16) \;.
\end{split}
\end{align}
Quite remarkably, the same hypergeometric functions also show up in the case of surface defects in the 6d $(2,0)$ theory and the $c_{mn}$ coefficients in that setup have a similar structure~\cite{Chen:2024orp}. 

\subsection{Evaluating the Mellin integral}\label{subsec:Mellin_residues}
Having obtained the general $c_{mn}$ coefficients from the $\log B\log^2D$ term, we next need to verify that in fact the full LLS in both channels is correctly reproduced. To this end, we explicitly evaluate the Mellin integral \eqref{eq:Mellin transform} by taking residues, giving rise to a series representation in small $B$ and $D$ of the position space correlator which is of the form
\begin{align}\label{eq:H2_BD_expansion}
    \mathcal{H}^{(2)}(z,\bar{z})=\sum_{i,j=0}^\infty \mathcal{R}_{i,j}\,B^{i}D^{4+2j}\;.
\end{align}
The coefficients $\mathcal{R}_{i,j}$ are simply the residues at $\delta=-i$ and $\gamma=3+2j$, i.e.
\begin{equation}\label{eq:R_ij}
    \mathcal{R}_{i,j} \equiv -\text{Res}_{\delta=-i,\,\gamma=3+2j} \left\{\widetilde{\M}^{(2)}_{22}(\delta, \gamma)\, \Gamma(\delta)\Gamma(\gamma-\delta+1)\Gamma\!\left(\frac{3-\gamma}{2}\right)^2 \right\}\,,
\end{equation}
where we recall that the one-loop Mellin amplitude $\widetilde{\M}^{(2)}_{22}(\delta, \gamma)$ is given as a double-infinite sum over simultaneous simple poles labeled by $m,\,n$.

Let us first focus on extracting the full LLS in both channels, corresponding to the blue and green shaded areas in Figure \ref{fig:log-power-table}. For the bulk channel LLS, this amounts to considering all double poles in $\delta$, leading to a single infinite sum over $m$ for each fixed value of $n$. Similarly, for the defect channel LLS one needs to take triple poles in $\gamma$, and one is left with an infinite sum over $n$ with $m$ fixed. In both cases these sums are convergent, and one straightforwardly obtains the full LLS as a series in small $B$ and $D$. While the $\log B\,\log^2D$ terms agree by construction, we find that in addition also the $\log B\,\log D$, $\log B$, and the $\log^2D$ terms are in perfect agreement with the explicit OPE resummations from Section~\ref{Subsec:resumedLLS}. We thus conclude that our ansatz \eqref{eq:MellinAnsatz} indeed contains all the needed singular terms and no additional single poles are present.

On the other hand, calculating the subleading logarithmic terms (corresponding to the grey shaded area in Figure \ref{fig:log-power-table}) via \eqref{eq:R_ij} leads to a double infinite sum over $m$ and $n$ which does not converge. In order to obtain a finite answer, we thus need to regularize this sum. This can be achieved by introducing a \textit{regularized} Mellin amplitude
\begin{equation}\label{eq:M2_reg}
    \widetilde{\mathcal{M}}^{(2)}_{22,\,{\rm reg}}(\delta, \gamma) = \sum_{m,n=0}^{\infty} \left(\frac{c_{mn}}{(\delta + n)(\gamma - 3 - 2m)} +d_{mn}\right)+C\,,
\end{equation}
with some suitable regulator $d_{mn}$ that is independent of $\delta$ and $\gamma$. We choose this regulator such that it cancels the divergence in the residue evaluation \eqref{eq:R_ij}. A convenient choice is given by\footnote{It turns out that this regulator also renders the Mellin amplitude \textit{itself} finite, not just the sum over residues. This can be seen by reorganizing the summation over $m$ and $n$. Since the divergence occurs when both $m$ and $n$ are large, we take $m=rt$ and $n=r(1-t)$ such that $m+n=r$. For large $r$, the summand of the unregularized Mellin amplitude behaves as $1/r$, signaling the divergence. Adding the regulator $d_{mn}$ removes the leading $1/r$ term and the regularized Mellin amplitude goes as $1/r^2$ instead, such that the remaining sum over $r$ is convergent.}
\begin{equation}
    d_{mn}=-\frac{3\left(m^2+n^2+2mn-m-n-9\right)}{4(m+n+2)^2(m+n+3)^2}\;.
\end{equation}
In the above regularized Mellin amplitude, we have also included an extra shift by a constant $C$ to account for the arbitrariness in the choice of regulator. It is useful to write this ambiguity as
\begin{equation}
   C=c_0-\frac{3\pi^2}{8}\;.
\end{equation}
Being a constant, this term corresponds to a two-point defect contact Witten diagram and has a natural interpretation as a one-loop counter-term ambiguity.

Using the regularized Mellin amplitude \eqref{eq:M2_reg} in the residue evaluation then leads to finite answers for all $\mathcal{R}_{i,j}$. For instance, for $i=j=0$ we have
\begin{align}
\begin{split}
    \mathcal{R}_{0,0}&=-\frac{2}{3}\Big[2\pi^2\log D+12\log B\,\log^2D-3\pi^2+18\log B\,\log D\\
    &\qquad\qquad\qquad\qquad\qquad\quad+(36c_0-120)\log D+66c_0-153\Big]\,.
\end{split}
\end{align}
Nevertheless, the computation of these residues for general $(i,j)$ remains technically difficult. We discuss further details of these calculations in Appendix~\ref{appendix:Residue Results} where we also include more explicit examples for $\mathcal{R}_{i,j}$ for other values of $(i,j)$. These result are relevant for the comparison of the Mellin space result with the corresponding position space representation, which we construct via a complementary position space bootstrap approach in Section \ref{Sec:positionspace}.

\subsection{Flat-space limit}

The Mellin amplitude also contains information about the flat-space amplitude, which is accessed via the regime of large Mellin variables $\delta$ and $\gamma$. The consistency with the flat-space result provides a further independent test of the exact formula \eqref{eq:generalCmn} and further constrains the possible regular term ambiguities. 

\vspace{0.4cm}

\noindent {\bf Flat-space formula for the reduced Mellin amplitude.} The flat-space limit of the defect Mellin amplitude can be extracted using the  formula of \cite{Penedones:2010ue,Alday:2024srr}. Here we will use only the structural part. For two scalar operators of dimensions $\Delta_1$ and $\Delta_2$ in the presence of a $p$-dimensional defect, the large-radius limit of the defect Mellin amplitude is obtained from the flat-space form factor by the Borel transform
\begin{equation}\label{Mellinfslimitf}
\mathcal M(\delta,\gamma)
\underset{R\rightarrow\infty}{\sim}\;
{\rm const.}\,\int_0^\infty d\beta\,\beta^{\frac{\Delta_1+\Delta_2-p}{2}-1}e^{-\beta}\mathcal A\left(S=-\frac{\beta\delta}{R^2}, Q=\frac{\beta\gamma}{R^2}\right)\;.
\end{equation}
Here $R$ is the AdS radius and $S,Q$ are the two flat-space kinematic invariants. We will not attempt to fix the overall normalization of this map and we will only compare the nontrivial dependence on the kinematic variables.

For the $k=2$ giant-graviton reduced correlator, the reduced external fields behave as effective scalars of dimensions $\Delta_1^{\rm red}=\Delta_2^{\rm red}=4$, while the defect dimension is $p=0$. Thus we have
\begin{equation}
\widetilde{\mathcal M}_{22}(\delta,\gamma)
\propto
\int_0^\infty d\beta\,\beta^3e^{-\beta}\mathcal A_{\rm red}\left(S=-\frac{\beta\delta}{R^2}, Q=\frac{\beta\gamma}{R^2}\right)\;,\qquad R\rightarrow\infty.
\end{equation}
If the reduced flat-space form factor is homogeneous of degree $r_L$,
\begin{equation}
\mathcal A_{\rm red}^{(L)}(\lambda S,\lambda Q)
=
\lambda^{r_L}\mathcal A_{\rm red}^{(L)}(S,Q)\;,
\end{equation}
then the Borel integral can be performed explicitly and only produces an overall Gamma-function factor,
\begin{equation}
\widetilde{\mathcal M}_{22}^{(L)}(\delta,\gamma)
\sim
\Gamma(4+r_L)\mathcal A_{\rm red}^{(L)}\left(-\frac{\delta}{R^2},\frac{\gamma}{R^2}\right)\;.
\end{equation}
In what follows we work with dimensionless flat-space invariants, obtained by absorbing the appropriate powers of the AdS radius $R$ into $S$ and $Q$. 

At tree level, the flat-space limit of giant-graviton correlators has already been matched to the corresponding flat-space form factor in \cite{Chen:2025cod}. In particular, after stripping off the universal supersymmetric kinematic factor, the tree-level reduced form factor behaves as
\begin{equation}
\mathcal A_{\rm red}^{(1)}(S,Q)
\propto
\frac{1}{SQ}\;.
\end{equation}
Under the flat-space identification $S\sim -\delta$ and $Q\sim \gamma$, this correctly reproduces the large Mellin-variable behavior of the tree-level reduced Mellin amplitude,
\begin{equation}
\widetilde{\mathcal M}_{22}^{(1)}(\delta,\gamma)
\underset{\delta,\gamma\rightarrow\infty}{\sim}
\frac{1}{\delta\gamma}\;,
\end{equation}
up to an overall normalization.

\vspace{0.4cm}

\noindent {\bf One-loop level.} The logarithmically divergent one-loop flat-space integral is homogeneous of degree zero, up to polynomial ambiguities, and should also reproduce the same kinematic dependence as the reduced Mellin amplitude at the flat-space limit. 
In the representation \eqref{eq:MellinAnsatz}, taking $\delta$ and $\gamma$ large probes the part of the double sum for which the summation variables $m$ and $n$ are themselves large. 
Therefore, in order to identify the flat-space limit of the one-loop correlator, one should determine the asymptotic behavior of the Mellin coefficients $c_{mn}$ in the simultaneous large-$m$, large-$n$ limit. We accordingly consider the scaling limit
\begin{equation}
m=\Lambda x\;,
\qquad
n=\Lambda y\;,
\qquad
\Lambda\to\infty \;,
\end{equation}
with $x$ and $y$ fixed.

Although the exact expression \eqref{eq:generalCmn} is compact, its large-$\Lambda$ behavior is obscured by substantial cancellations among the different hypergeometric contributions. 
After reorganizing the answer into a numerically stable form, we find that the coefficients approach 
\begin{equation}
c_{mn}
\longrightarrow
C(x,y)
=
\frac{12x\sqrt y}{(4x+y)^{3/2}}\; .
\end{equation}
In this scaling regime, the discrete double sum in the Mellin amplitude is approximated by the integral
\begin{equation}
I(\delta,\gamma)
=
\int_0^\infty dx \int_0^\infty dy\,
\frac{12x\sqrt y}{(4x+y)^{3/2}(\delta+y)(\gamma-2x)} \;.
\end{equation}
Although this integral is divergent, its physically meaningful content can be extracted by acting with suitable derivatives with respect to the kinematic variables. 
This is analogous to the treatment of flat-space loop integrals, where divergent polynomial ambiguities are removed by differential operators.

A first useful relation follows from a simple scaling argument and gives
\begin{equation}
\left(\delta\partial_\delta+\gamma\partial_\gamma\right)I(\delta,\gamma)=3 \;.
\end{equation}
As a consequence, the answer is of the form
\begin{equation}
I(\delta,\gamma)=3\log\delta+F(\rho)\;,
\qquad
\rho=\frac{\gamma}{\gamma-\delta} \;,
\end{equation}
where $F(\rho)$ depends only on the scale-invariant ratio $\rho$.

A second independent relation is obtained by differentiating with respect to $\gamma$. This yields a convergent integral and determines a first-order differential equation for the scale-invariant part $F(\rho)$, 
\begin{equation}
\partial_\gamma I(\delta,\gamma)
=
-\frac{(1-\rho)^2}{\delta}\,J(\rho)\;,
\end{equation}
where $J(\rho)\equiv F'(\rho)$. Introducing the variable $u=\sqrt{\frac{1+\rho}{1-\rho}}$, we have
\begin{equation}
\begin{split}
J(\rho)=\frac{3(1+u^2)^2}{16u^5}\bigg\{&8u^3-12u\log\!\left(\frac{4}{u^2-1}\right)-(u^2-3)\left[2\pi^2+\log^2(u^2-1)\right.
\\
&\left.-4\log^2(u+1)+4\log2\log\!\left(\frac{u+1}{u-1}\right)-4\operatorname{Li}_2\!\left(\frac{2}{u+1}\right)\right]\bigg\}\;.
\end{split}
\end{equation}
Integrating this first-order equation and expressing the result in $u$ gives  
\begin{equation}
\begin{split}
&F(\rho) = \frac{1}{4u^3} \bigg\{  12u \Big[ 2\log 2 + (u^2-1) \log(u^2-1) \Big] + \pi^2 \left( u^3 + 6u^2 - 6 \right) 
\\
& + 3(u^2-1) \left[ \log^2(u^2-1) - 4\log^2(u+1) + 4\log 2 \log \left( \frac{u+1}{u-1} \right) - 4\operatorname{Li}_2\left( \frac{2}{u+1} \right) \right] \bigg\}+C_{\rm flat}\;.
\end{split}
\end{equation}
The additive constant $C_{\rm flat}$ reflects the ambiguity in choosing the finite part of the logarithmically divergent flat-space integral and is not fixed by the first-order equation for $F(\rho)$. The complexity of this result, involving dilogarithms of weight 2, is consistent with the transcendental structure expected for one-loop holographic correlators.

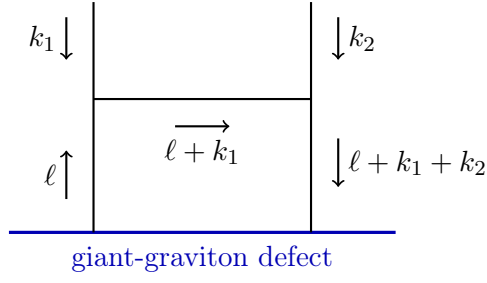
\begin{figure}[htbp]
    \centering
    \begin{tikzpicture}[x=0.8cm,y=0.8cm]
        \draw[very thick,blue!70!black] (-3.2,0) -- (3.2,0);

        % H-shaped propagators
        \draw[thick] (-1.8,3.8) -- (-1.8,0);
        \draw[thick] ( 1.8,3.8) -- ( 1.8,0);
        \draw[thick] (-1.8,2.2) -- ( 1.8,2.2);

        % External momenta
        \draw[->,thick] (-2.25,3.55) -- (-2.25,2.85);
        \node[left] at (-2.25,3.2) {$k_1$};

        \draw[->,thick] (2.25,3.55) -- (2.25,2.85);
        \node[right] at (2.25,3.2) {$k_2$};

        % Loop momentum 
        \draw[->,thick] (-2.25,0.55) -- (-2.25,1.35);
        \node[left] at (-2.25,0.95) {$\ell$};

        \draw[->,thick] (-0.45,1.75) -- (0.45,1.75);
        \node[below] at (0,1.70)
            {$\ell+k_{1}$};
            
        \draw[->,thick] (2.25,1.55) -- (2.25,0.75);
        \node[right] at (2.25,1.15)
            {$\ell+k_{1}+k_{2}$};

        \node[below,blue!70!black] at (0,-0.05)
            {giant-graviton defect};
    \end{tikzpicture}

    \caption{Flat-space one-loop Feynman diagram for two massless
    particles scattering off the giant-graviton defect. }
    \label{fig:flat-space-loop}
\end{figure}

What is important is that exactly the same structure arises from the corresponding flat-space one-loop Feynman integral. For the 1-to-1 scattering of massless gravitons with a giant graviton defect, the process includes in particular the Feynman diagram shown in Figure~\ref{fig:flat-space-loop}.\footnote{Note that the flat-space amplitude here is also in a reduced sense. The full amplitude before the reduction is made of more Feynman diagrams. Similar identifications of the  flat-space limit with a single Feynman diagram have also been made in defect-free cases for gravitons \cite{Alday:2018kkw,Alday:2019nin} and gluons \cite{Alday:2021ajh}, as well as in the defect case for a 't Hooft loop \cite{Alday:2024srr}. } 

 The relevant one-loop integral in  flat space is 
\begin{equation}
I_{\rm flat}
=
\int \frac{d^{d_{\perp}}\ell_{\perp}}{\pi^{\frac{d_{\perp}}{2}}}\,
\frac{1}{
\ell_{\perp}^2\big[(\ell_{\perp}+k_{1,\perp})^2+k_{1,\parallel}^2\big]
(\ell_{\perp}+k_{1,\perp}+k_{2,\perp})^2
} \;.
\end{equation}
Here the total spacetime dimension is ten and the giant graviton D3 brane is seen as a four-dimensional defect. Therefore, the transverse dimension is  $d_{\perp}=6$. The loop-momentum flowing in the transverse directions is denoted by $\ell_\perp$, while the external momenta are decomposed into components parallel and transverse to the defect, $k_i=(k_{i,\parallel},k_{i,\perp})$. 
Since the defect preserves translation invariance along its worldvolume, momentum is conserved in the parallel directions, which gives  $k_{1,\parallel}+k_{2,\parallel}=0$. Moreover, the on-shell condition implies $ k_{i,\perp}^2+k_{i,\parallel}^2=0$.
As a result, it is convenient to describe the scattering process using two flat-space kinematic invariants
\begin{equation}
Q=-k_{1,\parallel}^2=k_{1,\perp}^2=k_{2,\perp}^2\;,
\qquad
S=-\,k_{1}\!\cdot k_{2}\;,
\end{equation}
which we have already used in (\ref{Mellinfslimitf}).

Introducing the Schwinger parameters $a_1$, $a_2$, $a_3$ and performing the Gaussian integral over the loop momentum, one finds
\begin{equation}
I_{\rm flat}
=
\int_0^\infty \prod_{i=1}^3 d a_i
a_\Sigma^{\frac{-d_{\perp}}{2}}
\exp\!\left[
\frac{2a_1a_3}{a_\Sigma}(S+Q)
-\frac{2a_1a_3-a_2^2}{a_\Sigma}Q
\right],
\end{equation}
where $a_\Sigma=a_1+a_2+a_3$. 
We then pass to the Feynman parameters by setting $a_i=\beta_i t$ with the constraint $\sum_i\beta_i=1$, and use the dimensionless convention $S\to -\delta,
Q\to \gamma$. The integral becomes
\begin{equation}
I_{\rm flat}(\delta,\gamma)
=
\int_0^\infty dt\, t^{2-d_{\perp}/2}
\int_0^1 d\beta_1 d\beta_2 d\beta_3\,
\boldsymbol{\delta}\!\left(\sum_i\beta_i-1\right)e^{-tA},
\end{equation}
with
\begin{equation}
A=2\beta_1\beta_3\,\delta-\beta_2^2\gamma \;.
\end{equation}

For $d_{\perp}=6$, we have $t^{2-d_{\perp}/2}=t^{-1}$, so the integral is logarithmically divergent as $t\to 0$. 
As in the Mellin analysis, the divergent part can be removed by differentiating with respect to the kinematic variables. Acting with the Euler operator $D_E=\delta\partial_\delta+\gamma\partial_\gamma$, or with $\partial_\gamma$, we can reduce $I_{\rm flat}$ to a convergent integral. The resulting differential equations agree with those satisfied by the Mellin amplitude integral up to an overall factor, namely we have
\begin{equation}
-6\left(\delta\partial_\delta+\gamma\partial_\gamma\right)I_{\rm flat}(\delta,\gamma)=3 \;,
\end{equation}
\begin{equation}
-6\, \partial_\gamma I_{\rm flat}(\delta,\gamma)=-\frac{(1-\rho)^2}{\delta}\,J(\rho)\;.
\end{equation}
Consequently, the flat-space Feynman integral and the Mellin amplitude integral have the same nontrivial functional dependence on the kinematics, differing at most by the polynomial ambiguities of logarithmically divergent integrals. This provides a test of the exact Mellin coefficients \eqref{eq:generalCmn} and confirms that the Mellin representation correctly captures the behavior at the flat-space limit.

\section{The one-loop correlator in position space}\label{Sec:positionspace}

We now turn to the position space representation of the one-loop correlator. It turns out that, despite the appearance of infinite sums over hypergeometric $_3F_2$'s in the Mellin amplitude, the correlator admits an explicit position space representation in terms of a surprisingly simple set of transcendental functions. In fact, the form of the leading-logarithmic singularities obtained in Section \ref{section:unitarity_method} already reveals a part of the singularity structure which the position space correlator must possess. This, together with compatibility with crossing symmetry, leads us to consider an ansatz for the one-loop correlator in terms of a minimal basis of transcendental functions. Imposing various physically motivated constraints on this ansatz fixes all but a few parameters, which we then determine by comparing to the Mellin space result of Section \ref{Sec:Mellinspace}. At the same time, as this comparison is highly over-constrained, it provides a nontrivial consistency check on the validity of the position space result.

\subsection{The basis of transcendental functions}\label{subsec:basis}
Let us start by motivating a suitable set of basis functions which describe the one-loop correlator $\H^{(2)}(z,\zb)$ in position space. In order to reproduce the structure of logarithmic discontinuities, we expect the basis elements to be transcendental functions with at most $\log U$ behavior for small $U$, and at most $\log^2V$ behavior for small $V$. Moreover, based on the explicit results for the LLS given in eqs. \eqref{eq:defect resum} and \eqref{eq:bulk resum}, we deduce that their transcendental weight can be at most three. In addition, due to the symmetry of the correlator \eqref{eq:H_crossing}, the basis of functions needs to be closed under $z\mapsto z'$ transformations (corresponding to $(U,V)\mapsto(U/V,1/V)$).

Altogether, these requirements are quite constraining and lead us to consider the following seven basis functions $\Q_i(z,\zb)$:
\begin{align}\label{eq:basis8}
\begin{split}
	\text{weight 3}^-:&\qquad\quad\Q_{1}=\log V~\phi^{(1)}(z,\zb)\,,\\
	\text{weight 2}^-:&\qquad\quad\Q_{2}=\phi^{(1)}(x,\xb)\,,\\
	\text{weight 2}^+:&\qquad~\,\Q_{3,4}=\log U \,\log V\,,\,\log^2V\,,\\
	\text{weight 1}^+:&\qquad~\,\Q_{5,6}=\log U\,,\,\log V\,,\\
	\text{weight 0}^+:&\qquad\quad\Q_{7}=1\,,
\end{split}
\end{align}
which we organized according to their transcendental weight and parity, even $(+)$ or odd $(-)$, under $z\leftrightarrow\zb$ exchange. Here $\phi^{(1)}(z,\zb)$ denotes the standard one-loop box integral, whose explicit form is given by\footnote{Note that this function is fully antisymmetric under crossing transformations, i.e.
\begin{align}
	\phi^{(1)}(z,\zb) = -\phi^{(1)}(z',\zb') = -\phi^{(1)}(1/z,1/\zb) = -\phi^{(1)}(1-z,1-\zb)\,.
\end{align}}
\begin{align}
	\phi^{(1)}(z,\zb) = \log U\,\big[\log(1-z)-\log(1-\zb)\big]+2\big[\Li_2(z)-\Li_2(\zb)\big]\,.
\end{align}
The above set of functions is in fact a subset of the transcendental basis used to describe defect-free four-point correlators at one-loop order, see e.g. \cite{Aprile:2019rep}. 

However, it turns out that the seven functions given above are not sufficient to fully capture the leading log resummations simultaneously in both channels. This can be remedied by including an additional, eighth basis function given by\footnote{We note that the new ingredients in this function, $\Li_2(V)$ and $\log(1-V)$, naturally appear in one-loop bubble Witten diagrams on the defect. See also footnote~\ref{foot:Q8} for a comment on the concrete computational reasons for the presence of the function $\Q_8(z,\zb)$.}
\begin{align}
	\Q_8 = \Li_2(V)-\frac{1}{4}\log^2 V+\log(1-V)\log V-\frac{\pi^2}{6}\,,
\end{align}
which according to the above classification should be part of the weight $2^+$ row. Moreover, one can check using the inversion formula of the dilogarithm that this function satisfies $\Q_8(z,\zb)=-\Q_8(z',\zb')$, and therefore does not mix with the other basis elements under crossing.

In modern parlance, we thus conclude that the transcendental functions needed to describe the one-loop defect correlator in position space are built from the alphabet of letters $\{z,\zb,1-z,1-\zb,V,1-V\}$. In contrast, similar bootstrap calculations have shown that one-loop four-point functions in the defect-free case contain functions built from the simpler set of letters $\{z,\zb,1-z,1-\zb\}$, although there the maximal transcendental weight goes up to four.

\subsection{The one-loop bootstrap in position space}\label{subsec:pos_space_bootstrap}
We are now equipped to construct an ansatz for the one-loop correlator $\H^{(2)}(z,\zb)$. In order to reproduce the leading log resummations derived in Section \ref{section:unitarity_method}, the transcendental functions $\Q_i(z,\zb)$ need to be multiplied by certain coefficient functions. This motivates an ansatz of the form
\begin{align}\label{eq:H_ansatz}
	\H^{(2)}_{22}(z,\zb) = \frac{1}{(1-V)^7}\sum_{i=1}^8\frac{p_i(z,\zb)}{(z-\zb)^{d_\pm}}\,\Q_i(z,\zb)\,,
\end{align}
where the $p_i$ are symmetric polynomials given by $p_i(z,\zb)=\sum_{n,m}^{14} a^{(i)}_{n,m}(z^n\zb^m+z^m\zb^n)$ with unfixed numerical parameters $a_{n,m}^{(i)}$. The denominator powers $d_+=6$ and $d_-=7$ are correlated with the parity of each basis function $\Q_i(z,\zb)$ under $z\leftrightarrow\zb$ exchange, ensuring that overall the ansatz is symmetric as required.

Taking this ansatz as a starting point, we proceed to constrain the free parameters $a^{(i)}_{n,m}$ by imposing the following bootstrap conditions:
\begin{itemize}\itemsep=0pt
\item[(1)]\textbf{Crossing symmetry:} the correlator has to satisfy \eqref{eq:H_crossing}, i.e., invariance under $(z,\zb)\mapsto(z',\zb')$ transformation. 

\item[(2)]\textbf{Absence of unphysical poles:} the poles at $z=\zb$ which are present in the ansatz \eqref{eq:H_ansatz} need to be canceled as such poles are unphysical for a correlator in Euclidean signature.

\item[(3)]\textbf{Matching the LLS:} the leading log discontinuities in the two channels have to match the resummations given in eqs. \eqref{eq:defect resum} and \eqref{eq:bulk resum}.\footnote{\label{foot:Q8}
It is this step where the necessity to include the eighth basis element $\Q_8(z,\zb)$ in our ansatz becomes apparent: without it, it is not possible to simultaneously match both the bulk- and defect-channel leading logs. The discrepancy is however restricted to the $\log^2V$ terms with rational coefficients which are function of $V$ only, motivating the need for an additional function of weight 2 which depends only on $V$.}

\item[(4)]\textbf{Absence of below-threshold contributions:} there should be no contributions to the one-loop correlator below the two-particle threshold in both OPE channels. In other words, the leading terms at small $(z,\zb)$ and $(x,\xb)$ should be of order $z^0\zb^0$ and $x^2\xb^2$, respectively.\footnote{In contrast, the expansions of tree-level correlator $\H_{22}^{(1)}(z,\zb)$ start at order $z^{-1}\zb^{-1}$ and $x^1\xb^1$. This condition is therefore equivalent to excluding any poles below the two-particle threshold in the Mellin space ansatz \eqref{eq:MellinAnsatz}, see also footnote \ref{foot:tree-level_amb}. 
}
\end{itemize}
Implementing this four-step algorithm determines all but 7 parameters. We thus arrive at a consistent result for the one-loop function $\H^{(2)}_{22}(z,\zb)$, which however still contains 7 ambiguities. Upon closer inspection, it turns out that these ambiguities are all `tree-like' functions, meaning they contribute only to the basis elements $\Q_6=\log V$ and $\Q_7=1$. Moreover, their denominator contains only powers of $(1-V)$, while the factors of $(z-\zb)$ are absent. One particular linear combination of these ambiguities corresponds to a contact Witten diagram -- the function corresponding to a constant Mellin amplitude. This is the expected one-loop counter-term ambiguity which is also present in the one-loop Mellin amplitude, see Section \ref{subsec:Mellin_residues}.

On the other hand, the six other ambiguities do not seem to have an obvious physical interpretation. It is also not immediately clear how to fix these ambiguities within the position space bootstrap algorithm only. This is in contrast to the analogous bootstrap approach to defect-free four-point functions. There, one also finds that certain tree-like ambiguities are left unfixed, but in all known cases these are in one-to-one correspondence with one-loop counter-terms which are also present in the Mellin amplitude. Besides those, there are no further unfixed parameters left.

At least at a computational level, the reason for the presence of these additional ambiguities is clear. The reason lies in the denominator powers of $(1-V)$ in \eqref{eq:H_ansatz}, which is a qualitatively new feature not present in the known defect-free cases. This renders the pole-cancellation condition (the constraint (2) above) less constraining. Indeed, as mentioned above, the ambiguities have no $(z-\zb)$ denominator and hence they are unconstrained by the $z=\zb$ pole cancellation.

Since we do not have an independent way of determining these remaining six free parameters, we shall proceed to fix them by matching against the Mellin space result obtained in Section~\ref{Sec:Mellinspace}, which does not suffer from these additional ambiguities.\\

\noindent{\bf Matching against Mellin space.} The comparison with the Mellin space result is most easily done in a series expansion around small $(z,1-\zb)$, corresponding to the expansion in small $(B,D)$ described in Section \ref{subsec:Mellin_residues}, see eq. \eqref{eq:H2_BD_expansion}. This determines six of the unfixed parameters, and relates the seventh one to the expected one-loop contact-term ambiguity parametrized by $c_0$. We have performed this matching using the explicit data for the seven residue contributions
$\mathcal{R}_{i,j}$ with
$(i,j)=(0,0),\,(1,0),\,(2,0),\,(3,0),\,(4,0),\,(0,1),\,(1,1)$,
quoted in eqs.~\eqref{eq:R00}--\eqref{eq:R11}.

We emphasize that we have considered more terms than we have unknowns, meaning that the perfect agreement constitutes a nontrivial consistency check of the position space result, and in particular of the postulated basis of transcendental functions $\Q_i(z,\zb)$.

\subsection{The one-loop correlator $\H_{22}^{(2)}(z,\zb)$}\label{subsec:pos_space_result}
After having fixed the parameters in the one-loop ansatz \eqref{eq:H_ansatz}, we find that the position space correlator can be written as
\begin{align}\label{eq:H2_result}
	\H^{(2)}_{22}(z,\zb) = \dfourhat\P^{(2)}(z,\zb) + \frac12\H^{(1)}(z,\zb)\,.
\end{align}
Here $\H^{(1)}(z,\zb)$ is the tree-level correlator quoted in \eqref{eq:H1}, and $\dfourhat$ is a fourth-order differential operator defined as
\begin{align}\label{eq:dfour}
	\dfourhat = \frac{V^2}{z-\zb}\Big(\partial_{\zb}-\partial_z+\partial_z\partial_{\zb}(z-\zb)\Big)\partial_z\partial_{\zb}\,,
\end{align}
which as explained in \cite{Chen:2026ium} is natural to consider from the defect channel OPE point of view.\footnote{In particular, the differential operator $\dfourhat$ has the property that its eigenfunctions are the defect-channel long superconformal blocks (in the R-symmetry singlet channel). The corresponding eigenvalues are certain fourth-order polynomials which coincide with the numerator of the tree-level anomalous dimensions $\widehat{\gamma}^{(1)}$. See Section 6 of \cite{Chen:2026ium} for more details and for the generalization to all R-symmetry channels. Therefore, with the defect channel OPE in mind, acting with $\dfourhat$ effectively inserts a part of the tree-level anomalous dimension. This motivates us to write the one-loop correlator as in \eqref{eq:H2_result}.
}
It furthermore has the nice property that it respects the symmetry \eqref{eq:H_crossing} of the correlator, namely it is invariant under the $z\mapsto z'$ transformation. This differential operator acts on a simpler `pre-correlator' $\P^{(2)}(z,\zb)$ which can be expressed in the basis of transcendental functions $\Q_i(z,\zb)$ with coefficient functions $q_i(z,\zb)$, i.e.
\begin{align}
	\P^{(2)}(z,\zb)=\sum_{i=1}^8 q_i(z,\zb)\,\Q_i(z,\zb)\,,
\end{align}
with the explicit coefficient functions being given by
\begin{align}
\begin{split}
	q_1&=-\frac{U^2V^2}{2(z-\zb)^3(1-V)^3}\,,\\
	q_2&=-\frac{1}{8(z-\zb)^3(1-V)^2}\Big[U^3(1-V)^2+U^2(V+1)(V(3V-8)+3)\\
    &\qquad\qquad\qquad\qquad\qquad~+3U(V^2+1)(1-V)^2+(V+1)(V-1)^4\Big]\,,\\
	q_3&=\frac{U^2(V+1)((V-4)V+1)-2U(V^4-2V+1)+(V(V+2)-1)(V-1)^3}{16(z-\zb)^2(1-V)^3}\,,\\
	q_4&=-\frac{U^2(V+1)((V-4)V+1)+2U(-V^4+V^3+V-1)+(V+1)(V-1)^4}{8(z-\zb)^2(1-V)^3}\,,\\
	q_5&=\frac{U^2V}{2(z-\zb)^2(1-V)^2}\,,\\
	q_6&=-\frac{1}{48(z-\zb)^2(1-V)^3}\Big[U^2(7V^3-57V^2-33V+7)\\
    &\qquad\qquad\qquad\qquad\qquad\quad+(V-1)^2(7V^3-45V^2-45V+7)\\
	&\qquad\qquad\qquad\qquad\qquad\quad-U(14V^4-64V^3-204V^2-64V+14)\Big]\\
    &\qquad\qquad\qquad\qquad\qquad+\frac{V^3-3V^2-3 V+1}{6(1-V)^3}\,c_0\,,\\
	q_7&=\frac{7V}{12(1-V)^2}-\frac{2V}{3(1-V)^2}\,c_0\,,\\
	q_8&=\frac{(V+1)((V-4)V+1)}{4(1-V)^3}\,.
\end{split}
\end{align}
A few comments are in order. First, we note that the coefficient functions $q_i$ have reduced denominator powers of at most 3, which is because of the pulled out  fourth-order differential operator $\dfourhat$. Second, due to the invariance of $\dfourhat$, the pre-correlator $\P^{(2)}(z,\zb)$ enjoys the same $z\mapsto z'$ symmetry as the full correlator. Third, the counter-term ambiguity parametrized by $c_0$ can also be written as part of the pre-correlator where it contributes to coefficient functions $q_6$ and $q_7$, as expected from a tree-like function. Lastly, we emphasize that one should not read too much into the coefficient $\frac12$ of the tree-level term $\H^{(1)}$ in \eqref{eq:H2_result}. As mentioned in footnote \ref{foot:tree-level_amb}, an additional ambiguity that cannot be fixed without further input is the tree-level correlator itself. As such, the position space corelator as written in \eqref{eq:H2_result} simply matches the regularized Mellin amplitude \eqref{eq:M2_reg}, and the additional ambiguity of adding $\H^{(1)}$ is left implicit.

\subsection{Extracting the one-loop anomalous dimension}\label{subsec:gamma2}

In this subsection we extract the lowest-twist defect CFT data from the one-loop correlator $\mathcal H_{22}^{(2)}$. The reason we focus only on the family of twist-2 operators is that for higher twists there is operator mixing and one would need to consider one-loop correlators of higher KK-modes to resolve the mixing. We will start by recalling the lower-order data needed for the perturbative block expansion, and then use the complete position-space answer constructed above to calculate the one-loop anomalous dimension.

Our starting point is the defect channel OPE decomposition of the reduced correlator, which can be decomposed into long blocks $\widehat{g}_{\widehat{\Delta},s}$ with coefficients denoted by $\widehat A_{m,s}$. To order $1/c$ in the large-$c$ expansion, we then have
\begin{align}
\begin{split}
    \widehat A_{m,s}\,\widehat{g}_{\widehat{\Delta},s} &= \bigg\{\widehat A_{s}^{(0)}+\frac{1}{\sqrt{c}}\Big[\widehat A_{s}^{(1)}+\frac{1}{2}\widehat A_{s}^{(0)}\widehat\gamma^{(1)}_s\log V\Big]+\frac{1}{c}\Big[\widehat A_{s}^{(2)}\\
    &\qquad+\frac{1}{2}\big(\widehat A_{s}^{(1)}\widehat\gamma^{(1)}_s+\widehat A_{s}^{(0)}\widehat\gamma^{(2)}_s\big)\log V+\frac{1}{8}\widehat A_{s}^{(0)}(\widehat\gamma^{(1)}_s)^2\log^2V\Big]+\ldots\bigg\}\,\widehat g_{4+s,s}\;,
\end{split}
\end{align}
where on the RHS we dropped the label $m$ since we only focus on the contribution of double-particle operators of lowest twist corresponding to $m=0$.

In order to extract the one-loop anomalous dimensions $\widehat\gamma^{(2)}_s$, we first need to extract the relevant lower-order defect CFT data. The zeroth-order OPE coefficient and the tree-level anomalous dimension are obtained by specializing the decompositions used in Section 6 of \cite{Chen:2026ium} to the lowest-twist family $m=0$, yielding
\begin{equation}\label{eq:lowest-defect-a0-gamma1}
    \widehat A_s^{(0)}=s+1\;,\qquad
    \widehat\gamma_s^{(1)}=-\frac{4}{s+1}\;.
\end{equation}
Here $\widehat A_s^{(0)}$ comes from the free bulk-propagator term $G^{(0)}_{\rm bulk}$, while $\widehat\gamma_s^{(1)}$ is encoded in the $\log V$ coefficient of the tree-level correlator. 

It remains to determine the first correction to the OPE coefficient $\widehat A_s^{(1)}$, which can be extracted from the regular part of the tree-level correlator. Using the result for $\mathcal G_{22}^{({\rm tree})}$ from eqs. (5.15)-(5.16) of \cite{Chen:2026ium} and performing the defect channel superconformal block decomposition, we find that the twist-2 coefficient of the long blocks is simply given by\footnote{
For general twist and spin, the tree-level defect OPE data extracted from the $\llangle \mathcal{O}_2\mathcal{O}_2\rrangle$ correlator reads
\begin{align}
    \langle\widehat A^{(0)}_{m,s}\widehat\gamma^{(1)}_{m,s}\rangle = -2(m+1)(m+2)\,,\qquad\langle\widehat A^{(1)}_{m,s}\rangle = -(2m+3)\,,
\end{align}
where we recall that the angle brackets denote the averaging over the double-trace degeneracies. One then finds that this averaged defect OPE data obeys the familiar `derivative relation' \cite{Heemskerk:2009pn},
\begin{align}
    \langle\widehat A^{(1)}_{m,s}\rangle = \frac12\partial_m\,\langle\widehat A^{(0)}_{m,s}\widehat\gamma^{(1)}_{m,s}\rangle\,,
\end{align}
which was first observed to hold in the defect-free four-point function case.
}
\begin{equation}\label{eq:lowest-defect-a1}
    \widehat A_s^{(1)} = -3\;.
\end{equation}

Then, extracting the OPE data encoded in the $\log V$ coefficient of our one-loop correlator $\H_{22}^{(2)}$, we find
\begin{equation}
    \frac{1}{2}\big(\widehat A_{s}^{(1)}\widehat\gamma^{(1)}_s+\widehat A_{s}^{(0)}\widehat\gamma^{(2)}_s\big) =
    \begin{cases}\frac{191}{4}-12c_0, & s=0\;,\\[5pt]
    \displaystyle\frac{2\left(3s^3+13s^2-14s-18\right)}{s(s+1)^2(s+4)}, & s\geq 1\;.
    \end{cases}
\end{equation}
Together with the lower-order data \eqref{eq:lowest-defect-a0-gamma1} and \eqref{eq:lowest-defect-a1}, we finally isolate the one-loop anomalous dimensions:
\begin{equation}
    \widehat\gamma_s^{(2)}=
    \begin{cases}
    \displaystyle\frac{167}{2}-24c_0\;, & s=0\;,\\[6pt]
    \displaystyle-\frac{8(s^2+13s+9)}{s(s+1)^3(s+4)}\;, & s\geq 1\;.
    \end{cases}
\end{equation}
We note that this expression is analytic only for $s\geq1$, and the breakdown of analyticity in spin is manifest as the formula develops a pole at $s=0$. This is another manifestation of the fact that the one-loop correlator has a contact-term ambiguity, which contributes to the OPE data only for spin zero.

\section{Discussion and outlook}\label{Sec:discussion}

In this paper we constructed the first quantum correction to the correlator of two maximal giant gravitons and two stress-tensor multiplet operators in the supergravity limit of $\mathcal N=4$ SYM theory. By treating the two giant gravitons as a zero-dimensional defect, the problem of computing a four-point function reduces to the one of a defect two-point function. The main input was the tree-level giant graviton correlators of \cite{Chen:2025yxg,Chen:2026ium}, together with the defect version of the AdS unitarity method developed in \cite{Chen:2024orp}. In this way, a part of the one-loop answer is determined by gluing lower-order OPE data from which the full correlator can be reconstructed. Beyond the 6d surface-defect setup considered in \cite{Chen:2024orp}, our calculation is another example which provides further evidence for the general applicability of the defect two-point function unitarity method. In particular, it shows that this method continues to work in the limit of a zero-dimensional defect relevant for light-light-heavy-heavy correlators.

We obtained the explicit one-loop correlator in two complementary representations. In Mellin space, the reduced one-loop amplitude is captured by the simultaneous-pole ansatz \eqref{eq:MellinAnsatz}. The numerator coefficients $c_{mn}$ are fixed by matching the bulk- and defect-channel leading logarithmic singularities and admit the closed form \eqref{eq:generalCmn} in terms of hypergeometric $_3F_2$'s. Somewhat unexpectedly, the same type of hypergeometric building blocks also appeared in the surface-defect computation of \cite{Chen:2024orp}, which might suggest that this structure is not an accident of the present examples. The Mellin representation also makes the expected one-loop counter-term ambiguity manifest: it corresponds to the choice of regularization scheme for the divergent double sum in the Mellin amplitude. This ambiguity is simply a constant in Mellin space, i.e., a two-point defect contact Witten diagram.

We also constructed the same one-loop correction directly in position space. The leading logarithmic singularities strongly constrain the possible transcendental functions and lead to an ansatz of maximal transcendental weight three. In contrast to the analogous bootstrap approach to defect-free four-point functions, we found that not all parameters in the position-space ansatz can be fixed by the usual bootstrap conditions alone. The remaining freedom is constrained to tree-like defect dynamics, which we fixed by matching against the Mellin space result. The final answer can then be compactly written as
\begin{equation}\label{eq:H2_result2}
    \H^{(2)}_{22}(z,\zb)=\dfourhat\P^{(2)}(z,\zb)+\frac12\H^{(1)}_{22}(z,\zb)\,,
\end{equation}
where the fourth-order differential operator $\dfourhat$ introduced in \cite{Chen:2026ium} has a natural interpretation from the defect channel OPE point of view. A notable outcome of the position-space analysis is the alphabet of letters which appears in the transcendental functions. Besides the familiar defect-free letters $\{z,\zb,1-z,1-\zb\}$, the giant graviton defect correlator requires the additional letters $V$ and $1-V$. The complete set of letters relevant for the one-loop correction is therefore
\begin{equation}
    \{z,\zb,1-z,1-\zb,V,1-V\}\,.
\end{equation}
The extra letters enter through the new basis element $\Q_8$, which is needed in order to match the leading logarithms in the bulk and defect channels simultaneously. This gives a concrete position-space answer to the question of what function space is generated by one-loop defect unitarity in the giant graviton setup.

As a nontrivial consistency check, we studied the flat-space limit of the Mellin amplitude. The reduced Mellin amplitude correctly reproduces the functional dependence of the corresponding one-loop flat-space diagram for scattering off the giant graviton defect. 
Finally, we extracted the one-loop correction to the anomalous dimensions of the lowest-twist defect-channel operators. These operators describe bound states of a giant graviton and a light supergravity mode, and their one-loop anomalous dimensions give the quantum correction to the corresponding binding energies in AdS.

We conclude this discussion by mentioning several open questions and natural directions for future work.\vspace{0.3cm}

\noindent{\bf Ambiguities.} We have determined the one-loop correction up to two tree-level contributions, namely the contact-term ambiguity (parametrized by $c_0$) as well as the tree-level correlator $\H^{(1)}$ itself. The coefficient of the latter one can be fixed by a careful analysis of the protected contributions to the OPE at order $1/c$. This relies on the knowledge of the free theory correlator at that order, which however is a nontrivial calculation due to the presence of the two giant graviton operators \cite{Jiang:2019xdz}. The contact-term ambiguity can then be determined by comparing against the integrated giant graviton correlator recently studied in \cite{Brown:2024tru,Brown:2026dhy}.\vspace{0.3cm}

\noindent{\bf Higher-charge correlators and hidden symmetry.} The most immediate extension of our results is to compute one-loop correlators involving light operators of higher Kaluza-Klein charges. As shown in \cite{Chen:2025yxg,Chen:2026ium}, all tree-level correlators can be obtained from a single generating functional thanks to the existence of a defect version of the ten-dimensional hidden conformal symmetry which was first observed in defect-free $\frac{1}{2}$-BPS correlators \cite{Caron-Huot:2021usw}. At one-loop order, this hidden symmetry remains a useful organizational principle \cite{Aprile:2019rep,Alday:2019nin}, and it would be very interesting to see whether a similar story persists in the defect case. Higher charges would also allow us to probe the role of multiple R-symmetry channels, which are absent in the reduced correlator $\H_{22}$ studied here.\vspace{0.3cm}

\noindent{\bf Higher loops.} The compact form of the one-loop answer \eqref{eq:H2_result2} in terms of the fourth-order operator $\dfourhat$ suggests that higher loops may be more constrained than a direct Witten-diagram picture would indicate. Since $\dfourhat$ acts diagonally on defect-channel long blocks and inserts the numerator of the tree-level anomalous dimensions, it is tempting to look for a two-loop construction organized by an iterated action of this operator. This would parallel the simplifications observed in defect-free four-point functions at two loops \cite{Huang:2021xws,Drummond:2022dxw}. A two-loop calculation would also clarify whether the new position-space alphabet found here remains sufficient, or whether additional letters need to be included beyond one-loop order.\vspace{0.3cm}

\noindent{\bf Flat-space limit in position space.} The comparison to the flat-space form factor was performed in the Mellin space formulation. It would be useful to understand the same limit also from the complementary position-space perspective. This prescription has recently been explored in \cite{Chen:2025cod}, but so far has been applied only to tree-level defect correlators. Extending this analysis to loop-level would constitute an independent consistency check, both of our one-loop result and the flat-space limit prescription itself. Moreover, it would shed light on the role of the additional letters $V$ and $1-V$, and how they reorganize into the flat-space loop integral.\vspace{0.3cm}

\noindent{\bf Higher-derivative corrections.} Another natural problem is to include stringy and higher-derivative corrections. At tree level, higher-derivative contact terms are strongly constrained by the flat-space limit and constraints from supersymmetric localization, see e.g., \cite{Alday:2024srr} for a related defect two-point function setup. One can ask whether the one-loop unitarity method is constraining enough to bootstrap higher-derivative corrections, in analogy to the defect-free case where this approach has been successfully implemented to various orders in a small $\alpha'$ expansion in \cite{Alday:2018kkw,Drummond:2019hel,Drummond:2020uni,Aprile:2022tzr}. This would provide a new window into quantum corrections to D-brane effective actions in AdS$_5\times$S$^5$.\vspace{0.3cm}

\noindent{\bf Beyond maximal giant gravitons.} Finally, it would be interesting to consider quantum corrections to correlators of non-maximal and dual giant graviton operators. Such operators -- or more general heavy $\frac{1}{2}$-BPS backgrounds for that matter -- should still admit a defect description, but the tree-level correlators should first be obtained and the mixing problem need to be revisited. Understanding which parts of the unitarity construction survive in these more general backgrounds would help clarify how far the defect perspective on heavy-heavy-light-light correlators can be pushed. \vspace{0.5cm}

\acknowledgments
The work of J.C., X.D., and X.Z. is supported by the NSFC Grant No. 12275273 and funds from Chinese Academy of Sciences. The work of H.P. is supported in part by the FWO projects G003523N, G094523N, and G0E2723N, and the KU Leuven C1 project C16/25/01. This work is also supported by the NSFC Grant No. 12247103.

\appendix
\section{Explicit expressions of one-loop leading logs}\label{appendix:resum poly}

In this appendix, we provide the polynomial coefficients for the one-loop LLS in defect and bulk channels. 

The defect channel LLS can be written in the manifestly symmetric form
\begin{equation}
\mathcal{H}^{(2)}_{22}(x,\bar x)\big|_{\log^2 V}
=\frac{x^2\bar x^2}{(x-\bar x)^7(1-x\bar x)^7}
\Big[
P^{\rm def}_{0}(x,\bar x)
-
P^{\rm def}_{1}(x,\bar x)
\log\!\left(\frac{1-x}{1-\bar x}\right)
\Big]\;.
\end{equation}
Here $P^{\rm def}_{0}(x,\bar x)$ is antisymmetric under
$x\leftrightarrow \bar x$, while $P^{\rm def}_{1}(x,\bar x)$
is symmetric. The explicit polynomials are
\begin{equation}
\begin{split}
P^{\rm def}_{0}(x,\bar x)=&\,x\bar x\Big[
x^9\bar x^3(9+2\bar x+\bar x^2)
+x^8\bar x^2(117-74\bar x-82\bar x^2-18\bar x^3+28\bar x^4)\\
&+x^7\bar x(189-402\bar x-474\bar x^2+606\bar x^3+107\bar x^4-136\bar x^5)\\
&+x^6(45-242\bar x-616\bar x^2+1650\bar x^3+417\bar x^4-788\bar x^5+136\bar x^7-28\bar x^8)\\
&-x^5(4+263\bar x-1242\bar x^2-1085\bar x^3+3780\bar x^4-788\bar x^6+107\bar x^7-18\bar x^8+\bar x^9)\\
&+x^4(-6+132\bar x+61\bar x^2-2156\bar x^3+3780\bar x^5-417\bar x^6-606\bar x^7+82\bar x^8-2\bar x^9)\\
&+x^3(-12+102\bar x-316\bar x^2+2156\bar x^4-1085\bar x^5-1650\bar x^6+474\bar x^7+74\bar x^8-9\bar x^9)\\
&+x^2(-24\bar x+316\bar x^3-61\bar x^4-1242\bar x^5+616\bar x^6+402\bar x^7-117\bar x^8)\\
&+x\bar x^2(24-102\bar x-132\bar x^2+263\bar x^3+242\bar x^4-189\bar x^5)
+\bar x^3(12+6\bar x+4\bar x^2-45\bar x^3)
\Big]\;,
\end{split}
\end{equation}
and
\begin{equation}
\begin{split}
P^{\rm def}_{1}(x,\bar x)=&\,2\Big[
x^{10}\bar x^4
+2x^9\bar x^3(10-12\bar x-3\bar x^2+3\bar x^4)\\
&+3x^8\bar x^2(16-52\bar x+10\bar x^2+48\bar x^3-9\bar x^4-20\bar x^5+6\bar x^6)\\
&+2x^7\bar x(10-78\bar x+66\bar x^2+198\bar x^3-237\bar x^4+30\bar x^5+62\bar x^6-30\bar x^7+3\bar x^8)\\
&+x^6(1-24\bar x+30\bar x^2+396\bar x^3-540\bar x^4-360\bar x^5+338\bar x^6+60\bar x^7-27\bar x^8)\\
&-6x^5\bar x(1-24\bar x+79\bar x^2+60\bar x^3-260\bar x^4+60\bar x^5+79\bar x^6-24\bar x^7+\bar x^8)\\
&+x^4\bar x^2(-27+60\bar x+338\bar x^2-360\bar x^3-540\bar x^4+396\bar x^5+30\bar x^6-24\bar x^7+\bar x^8)\\
&+2x^3\bar x(3-30\bar x+62\bar x^2+30\bar x^3-237\bar x^4+198\bar x^5+66\bar x^6-78\bar x^7+10\bar x^8)\\
&+3x^2\bar x^2(6-20\bar x-9\bar x^2+48\bar x^3+10\bar x^4-52\bar x^5+16\bar x^6)\\
&+x(6\bar x^3-6\bar x^5-24\bar x^6+20\bar x^7)
+\bar x^6
\Big]\;.
\end{split}
\end{equation}

The bulk channel LLS can be written as
\begin{equation}
\begin{split}
\mathcal{H}^{(2)}_{22}(z,\bar z)\big|_{\log U}
=&\frac{(1-z)^2(1-\bar z)^2}{(z-\bar z)^7(z+\bar z-z\bar z)^7}
\Bigg[
P^{\rm bulk}_{0}(z,\bar z)
+P^{\rm bulk}_{1}(z,\bar z)\log(1-z)\\
&-P^{\rm bulk}_{1}(\bar z,z)\log(1-\bar z)
-2P^{\rm bulk}_{2}(z,\bar z)
\log\!\left(\frac{1-z}{1-\bar z}\right)
\log\!\big((1-z)(1-\bar z)\big)
\Bigg]\;.
\end{split}
\end{equation}
Here $P^{\rm bulk}_{0}(z,\bar z)$ is antisymmetric under
$z\leftrightarrow\bar z$, while $P^{\rm bulk}_{2}(z,\bar z)$
is symmetric. The explicit polynomials are
\begin{align}
P^{\rm bulk}_{0}(z,\bar z)=&
z^{10}\left(
6\bar z+41\bar z^2-202\bar z^3+311\bar z^4
-221\bar z^5+76\bar z^6-11\bar z^7
\right)
\nonumber\\
&+z^9\left(
-27\bar z^8+208\bar z^7-625\bar z^6+1158\bar z^5-1468\bar z^4
+1072\bar z^3-306\bar z^2-12\bar z
\right)
\nonumber\\
&+z^8\left(
27\bar z^9-680\bar z^7+1851\bar z^6-1752\bar z^5+950\bar z^4
-756\bar z^3+360\bar z^2
\right)
\nonumber\\
&+z^7\left(
11\bar z^{10}-208\bar z^9+680\bar z^8-2790\bar z^6
+2880\bar z^5-84\bar z^4-456\bar z^3
\right)
\nonumber\\
&+z^6\left(
-76\bar z^{10}+625\bar z^9-1851\bar z^8+2790\bar z^7
-2520\bar z^5+840\bar z^4
\right)
\nonumber\\
&+z^5\left(
221\bar z^{10}-1158\bar z^9+1752\bar z^8-2880\bar z^7
+2520\bar z^6
\right)
\nonumber\\
&+z^4\left(
-311\bar z^{10}+1468\bar z^9-950\bar z^8+84\bar z^7-840\bar z^6
\right)
\nonumber\\
&+z^3\left(
202\bar z^{10}-1072\bar z^9+756\bar z^8+456\bar z^7
\right)
\nonumber\\
&+z^2\left(
-41\bar z^{10}+306\bar z^9-360\bar z^8
\right)
+z\left(
12\bar z^9-6\bar z^{10}
\right)\;,
\end{align}
\begin{align}
P^{\rm bulk}_{1}(z,\bar z)=&
z^{10}\left(
3\bar z^7-21\bar z^6+63\bar z^5-87\bar z^4
+42\bar z^3+18\bar z^2-24\bar z+6
\right)
\nonumber\\
&+z^9\left(
27\bar z^8-172\bar z^7+427\bar z^6-570\bar z^5
+556\bar z^4-412\bar z^3+132\bar z^2+24\bar z-12
\right)
\nonumber\\
&+z^8\left(
27\bar z^9-336\bar z^8+1424\bar z^7-2887\bar z^6
+2712\bar z^5-1130\bar z^4+496\bar z^3-342\bar z^2+36\bar z
\right)
\nonumber\\
&+z^7\left(
3\bar z^{10}-228\bar z^9+1608\bar z^8-4648\bar z^7
+7878\bar z^6-7164\bar z^5+2128\bar z^4+312\bar z^3+120\bar z^2
\right)
\nonumber\\
&+z^6\left(
-23\bar z^{10}+779\bar z^9-4143\bar z^8+8172\bar z^7
-9180\bar z^6+8022\bar z^5-3294\bar z^4-396\bar z^3
\right)
\nonumber\\
&+z^5\left(
79\bar z^{10}-1370\bar z^9+6078\bar z^8-9564\bar z^7
+4998\bar z^6-1908\bar z^5+1872\bar z^4
\right)
\nonumber\\
&+z^4\left(
-155\bar z^{10}+1386\bar z^9-4626\bar z^8+7212\bar z^7
-2874\bar z^6-1236\bar z^5
\right)
\nonumber\\
&+z^3\left(
176\bar z^{10}-956\bar z^9+1452\bar z^8-2112\bar z^7+1704\bar z^6
\right)
\nonumber\\
&+z^2\left(
-106\bar z^{10}+518\bar z^9-216\bar z^8-324\bar z^7
\right)
+z\left(
26\bar z^{10}-156\bar z^9+156\bar z^8
\right)\;,
\end{align}
and
\begin{align}
P^{\rm bulk}_{2}(z,\bar z)=&
z^{10}\left(
\bar z^4-4\bar z^3+6\bar z^2-4\bar z+1
\right)
\nonumber\\
&+z^9\left(
6\bar z^7-42\bar z^6+120\bar z^5-166\bar z^4
+114\bar z^3-42\bar z^2+16\bar z-6
\right)
\nonumber\\
&+z^8\left(
18\bar z^8-138\bar z^7+435\bar z^6-810\bar z^5
+954\bar z^4-618\bar z^3+147\bar z^2+6\bar z+6
\right)
\nonumber\\
&+z^7\left(
6\bar z^9-138\bar z^8+748\bar z^7-1720\bar z^6
+2046\bar z^5-1598\bar z^4+980\bar z^3-300\bar z^2-24\bar z
\right)
\nonumber\\
&+z^6\left(
-42\bar z^9+435\bar z^8-1720\bar z^7+3390\bar z^6
-3138\bar z^5+1148\bar z^4-276\bar z^3+204\bar z^2
\right)
\nonumber\\
&+z^5\left(
120\bar z^9-810\bar z^8+2046\bar z^7-3138\bar z^6
+2856\bar z^5-864\bar z^4-216\bar z^3
\right)
\nonumber\\
&+z^4\left(
\bar z^{10}-166\bar z^9+954\bar z^8-1598\bar z^7
+1148\bar z^6-864\bar z^5+540\bar z^4
\right)
\nonumber\\
&+z^3\left(
-4\bar z^{10}+114\bar z^9-618\bar z^8+980\bar z^7
-276\bar z^6-216\bar z^5
\right)
\nonumber\\
&+z^2\left(
6\bar z^{10}-42\bar z^9+147\bar z^8-300\bar z^7+204\bar z^6
\right)
\nonumber\\
&+z\left(
-4\bar z^{10}+16\bar z^9+6\bar z^8-24\bar z^7
\right)
+\bar z^{10}-6\bar z^9+6\bar z^8.
\end{align}

\section{Details on the evaluation of Mellin space residues}\label{appendix:Residue Results}

Here we give more details on the evaluation of the Mellin-space residues used in Section~\ref{subsec:Mellin_residues}. For each pair $(i,j)$, the coefficient $\mathcal R_{i,j}$ is obtained by taking the poles at $\delta=-i$ and $\gamma=3+2j$ in the Mellin integral, i.e.
\begin{equation}
    \mathcal{R}_{i,j} \equiv -\text{Res}_{\delta=-i,\,\gamma=3+2j} \left\{\widetilde{\M}^{(2)}_{22,\,{\rm reg}}(\delta, \gamma)\, \Gamma(\delta)\Gamma(\gamma-\delta+1)\Gamma\!\left(\frac{3-\gamma}{2}\right)^2 \right\}\,,
\end{equation}
where we use the regularized one-loop Mellin amplitude introduced in Section~\ref{subsec:Mellin_residues}. 

The evaluation of the residue requires an infinite double summation over the indices $m$ and $n$. For fixed $(i,j)$, the terms with $m=j$ or $n=i$ are simple and can be evaluated directly. The main difficulty is the remaining part with $m\neq j$ and $n\neq i$, where the coefficients $c_{mn}$ involve generalized hypergeometric functions ${}_3F_2$ and a direct double summation is highly nontrivial. To resolve this, we adopt a strategy by first observing the form of the summand for fixed $i$ and $j$. For example, for the leading case $i=0, j=0$, when $m,n\neq0$, the summand is given by:
\begin{equation}
    \mathcal{R}_{0,0}^{(m,n)} = \frac{\sqrt{\pi}\,\Gamma(m)}{\Gamma(m + \frac{2 \lfloor n/2 \rfloor + 5}{2})} \left[ \frac{(3 + 11m)}{6m}  +  \log D \right]\sum_{q=0}^{\lfloor n/2 \rfloor + 1} \mathcal{C}_q\, m^q \,,
\end{equation}
where the coefficients $\mathcal{C}_q$ are constants. For higher $(i,j)$, the summand with $m\neq j$ and $n\neq i$ admits a similar simplified form.

The computation proceeds by first performing the infinite summation over $m$. Then, by analyzing the result across several values of $n$, we identify the patterns and subsequently perform the summation over $n$ to obtain the contribution to $\mathcal{R}_{i,j}$. Implementing this procedure and including the directly evaluated terms with $m=j$ or $n=i$, we obtain the following results for different values of $(i,j)$:
{\allowdisplaybreaks[4]
\begin{align}
\mathcal R_{0,0}
&=-\frac{2}{3}\Big[2\pi^2\log D+12\log B\,\log^2D-3\pi^2+18\log B\,\log D \notag\\*
&\quad+(36c_0-120)\log D+66c_0-153\Big]\,. \label{eq:R00}\\[6pt]
\mathcal R_{1,0}
&=\frac{1}{3}\Big[12\pi^2\log D+72\log B\,\log^2D-10\pi^2+132\log B\,\log D \notag\\*
&\quad+24\log^2D+(288c_0-822)\log D+600c_0-1413\Big]\,. \label{eq:R10}\\[6pt]
\mathcal R_{2,0}
&=\frac{1}{6}\Big[-48\pi^2\log D-288\log B\,\log^2D+58\pi^2-384\log B\,\log D \notag\\*
&\quad-168\log^2D+528\log B+(3822-1440c_0)\log D-3288c_0+7617\Big]\,. \label{eq:R20}\\[6pt]
\mathcal R_{3,0}
&=\frac{1}{9}\Big[120\pi^2\log D+720\log B\,\log^2D-203\pi^2+564\log B\,\log D \notag\\*
&\quad+564\log^2D-2817\log B+(4320c_0-11314)\log D+10584c_0-25227\Big]\,. \label{eq:R30}\\[6pt]
\mathcal R_{4,0}
&=\frac{1}{72}\Big[-1440\pi^2\log D-8640\log B\,\log^2D+3216\pi^2-1728\log B\,\log D \notag\\*
&\quad-8208\log^2D+53364\log B+(159756-60480c_0)\log D-156816c_0+388111\Big]\,. \label{eq:R40}\\[6pt]
\mathcal R_{0,1}
&=-\frac{2}{3}\Big[28\pi^2\log D+168\log B\,\log^2D-73\pi^2+204\log B\,\log D \notag\\*
&\quad-102\log^2D+18\log B+(720c_0-2598)\log D+924c_0-2154\Big]\,. \label{eq:R01}\\[6pt]
\mathcal R_{1,1}
&=\frac{2}{3}\Big[132\pi^2\log D+792\log B\,\log^2D-329\pi^2+1092\log B\,\log D \notag\\*
&\quad-678\log^2D+66\log B+(4320c_0-14778)\log D+6264c_0-15195\Big]\,. \label{eq:R11}
\end{align}
}
These results are the data required for matching the coefficients of the position space ansatz.

\bibliography{refs}
\bibliographystyle{utphys}
\end{document}